\documentclass[aps,prc,showkeys,twocolumn,nofootinbib]{revtex4}

\usepackage{graphicx}

\usepackage{isotope}
\usepackage{physics}

\def\SCNTHuizhou{$^{(1)}$Southern Center for Nuclear-Science Theory (SCNT), Institute of Modern Physics, Chinese Academy of Sciences, Huizhou 516000, China}
\def\KinrKyiv{$^{(2)}$Institute for Nuclear Research, National Academy of Sciences of Ukraine, Kyiv, 03680, Ukraine}
\def\HISTKLLanzhou{$^{(3)}$Heavy Ion Science and Technology Key Laboratory, Institute of Modern Physics, Chinese Academy of Sciences, Lanzhou 730000, China}
\def\UnivCASBeijing{$^{(4)}$School of Nuclear Sciences and Technology, University of Chinese Academy of Sciences, Beijing 101408, China}
\def\BitpKjiv{$^{(5)}$Bogolyubov Institute for Theoretical Physics, Metrolohichna str., 14b, Kyiv, 03143, Ukraine}

\begin{document}


\title{Microscopic study of nuclei synthesis in pycnonuclear reaction $\isotope[12]{C} + \isotope[12]{C}$ in neutron stars}
%
%
%

\author{S.~P.~Maydanyuk$^{(1,2)}$}\email{sergei.maydanyuk@impcas.ac.cn}
\author{Ju-Jun Xie$^{(1,3,4)}$}\email{xiejujun@impcas.ac.cn}
\author{V.~S.~Vasilevsky$^{5}$}\email{vsvasilevsky@gmail.com}
\author{K.~A.~Shaulskyi$^{(2)}$}\email{shaulskyi@kinr.kiev.ua} 


\affiliation{\SCNTHuizhou}
\affiliation{\KinrKyiv}
\affiliation{\HISTKLLanzhou}
\affiliation{\UnivCASBeijing}
\affiliation{\BitpKjiv}



\date{\small\today}

\begin{abstract}
\begin{description}

\item[Purpose]
To investigate synthesis of nuclei in pycnonuclear reactions in dense medium of neutron stars on the basis of understanding, how the compound nucleus is formed during collision of two closest nuclei. To implement microscopic formulation of nuclear interactions and fusion in pycnonuclear reactions in dense medium.

\item[Methods]
(1) Nuclei synthesis in pycnonuclear reaction in dense medium of neutron star is investigated in the folding approximation of the cluster model. 
(2) Formation of compound nucleus in dense medium is studied with high precision on the basis of the method of Multiple Internal Reflections. 


\item[Results]
(1) Parameters and wave functions of resonance states of \isotope[24]{Mg} are determiened by the interaction of two \isotope[12]{C} nucleis. 
%
(2) Clear maxima of probability of formation of compound nucleus in dense stellar medium are established.
Such a phenomenon has not been studied by other methods.
%
(3) Difference between quasibound energies 
for potential of Woods-Saxon type and folding potentials with 
the shell-model approximation for wave functions 
is essential. 
%
(4) Formation of the compound nucleus is much more probable in the quasibound states
than in states of zero-point vibrations 
for all studied potentials. 
%
(5) Only the first quasibound energies for
$\isotope[12]{C} + \isotope[12]{C}$ are smaller than the barrier maximums for all studied potentials. 
So, at the first quasibound energy the compound nuclear system has barrier which prevents its decay going through tunneling phenomenon.
This is the new excited nucleus \isotope[24]{Mg} synthesised in the neutron star.



\item[Conclusions]

Cluster approach with the folding potential provides significant modification of the picture of formation of compound nucleus,
which was previously obtained concerning to the potential of Woods-Saxon type.
The highest precision is provided by the folding potential, 
created by semi-realistic nucleon-nucleon potental and shell-model description of the internal structure 
of interacting  $p$-shell nuclei.
%
Synthesis of isotopes of Magnesium from isotopes of Carbon is essentially more probable in quasibound states
than in states of zero-point vibrations.
So, the quasibound states are highly important in understanding of synthesis of nuclei in dense medium of neutron stars.

\end{description}

\end{abstract}


\maketitle


\section{Introduction
\label{sec.introduction}}

Last years the neutron stars have been attracting more interest of researchers.
One of motivations for that is
neutron star is a good laboratory for deeper understanding of nuclear forces playing key role in structure of nuclei and nuclear reactions.
Special interest is attracted to explore nuclear reactions and properties of nuclei in conditions in stars
where densities of stellar medium is higher than nuclear saturation. Such a type of nuclear burning occurs in the cold and dense cores of white dwarfs~\cite{Salpeter_VanHorn.1969.AstrJ} and
crusts of neutron stars~\cite{Schramm.1990.AstrJ,Haensel_Zdunik.1990.AstroPhysJ,Haensel_Zdunik.2003.AstroPhysJ}. 
This phenomenon known as pycnonuclear reaction~\cite{Cameron.1959b.AstrJ}.
Here, nuclei are located at close distances between each others, can overlap and interact.
Pycnonuclear reactions are almost temperature independent and occur even at $T = 0$.
To study reactions with nuclei of different masses in the dense medium on safe basis, new special methods should be developed~\cite{Maydanyuk_Shaulskyi.2022.EPJA,Maydanyuk.2023.Universe}.



The first idea to study pycnonuclear reactions in stars was suggested by Gamov in 1938 \cite{Gamow.1939.PhysRev}
(see Ref.~\cite{Yakovlev.2005.EurPhysJA} on review of these reactions in compact stars).
Upon Gamow's request, Wildhack the first calculated rates of these reactions \cite{Wildhack.1940.PhysRev}. 
Salpeter and van Horn get strict formulation of estimation of pycnonuclar reaction rates
who also indicated three other important regimes on nuclear burning in dense matter~\cite{Salpeter.1969.AstroPhysJ}.
Later, pycnonuclear reactions have been studied in Refs.~\cite{Schramm_Koonin.1990.AstroPhysJ,Schramm_Koonin.1991.AstroPhysJ,Kitamura.1960.AstroPhysJ}.
Insight to this phenomenon was provided by Zel'dovich
who estimated zero-point energy as energy of the ground state of the harmonic oscillator potential energy formed 
close to saddle point 
between two close nuclei at lattice sites~\cite{Zeldovich.1965.AstrJ}.
Rates of pycnonuclear reactions have been estimated at such zero-point energies for a number of nuclei 
\cite{ShapiroTeukolsky.2004.book}.

Fusion is key process in pycnonuclear reactions. In this process new nucleus with larger mass is produced from two closest nuclei in lattice sites.
Fusion with isotopes of Carbon and nuclei with close masses have been intensively studied by different research groups
at energies of astrophysical interests~\cite{Yakovlev.2006.PRC,Kravchuk.2014.PRC}.
In Refs.~\cite{Afanasjev.2010.ADNDT,Singh.2019.NPA} $S$-factors were calculated for
946 fusion reactions including stable and neutron-rich isotopes of C, O, Ne, and Mg at energy in range from 2 to $\approx 18$-30~MeV.
Large collection of astrophysical $S$ factors and their compact representation for
isotopes of Be, B, C, N, O, F, Ne, Na, Mg, and Si
were presented in Ref.~\cite{Afanasjev.2012.PRC}.
Large database of $S$-factors was formed for about 5000 nonresonant fusion reactions.
However, those studies provide only partial understanding about nuclear reactions at low energies.
Actually, those investigations did not consider nuclear processes at condition of compact stars, 
where two nuclei located at close distances start to collide to each other.





Important issue in understanding of fusion is its high dependence on additional parameters
which appear in approaches outside the semiclassical approximations 
at energies close to barrier maxima~\cite{Maydanyuk.2023.Universe}.
Those parameters describe fusion process more dynamically
that was omitted in study indicated above.
Calculated cross sections of fusion are changed significantly 
if to vary these parameters. 
%
To initiate a process of collision between nuclei at close distances in the model, 
it needs to fix incident quantum flux of one nucleus on another nucleus located on lattice site without any reflectional fluxes.
But, interference between ingoing and outgoing fluxes exists.
It turns out that such an interference sometimes has larger role than ingoing flux or outgoing flux
at close distances between nuclei (i.e., in condition of neutron stars).
That phenomenon is studied by methods of quantum mechanics with high precision. 



In Ref.~\cite{Maydanyuk_Shaulskyi.2022.EPJA} an idea was proposed to study and estimate synthesis on more heavy nuclei in the pycnonuclear reactions 
on the basis of understanding of compound nucleus which can be formed 
during collision of incident nucleus on the closest another nucleus located in the lattice site. 
As was found, energies of states with the most probable formation of compound nucleus (called as quasibound states) in this reaction are different 
from energies of states of zero-point vibrations.
Note that probabilities of formation of compound nucleus in these quasibound states are 
much larger than in states of zero-point vibrations~\cite{Maydanyuk.2023.Universe}.
In particular, for $\isotope[12]{C} + \isotope[12]{C} \to \isotope[24]{Mg}$
ratio between probabilities of formation of compound nucleus for quasibound state and state of zero-point vibration is
approximately proportional to the corresponding penetrabilities of barrier that is
$T^{(\rm quasibound)} (E_{\rm quasibound} = 5.00~MeV) / T^{(\rm zero-mode)} (E_{\rm zero-mode} = 0.58~MeV) = 
6.811 \times 10^{+30}$ (see Eqs.~(45)--(46) in Ref.~\cite{Maydanyuk_Shaulskyi.2022.EPJA},
see also references for method, tests, demonstrations for different nuclear reactions, etc.).
Note that this approach provides also high precision tests for checking calculations which is fulfilled with accuracy up $10^{-14}$
(there is no other approaches providing such accuracy in estimation of penetrabilities of barriers in nuclear physics).
Nuclear reaction goes on the most probable channel.
Thus, this result leads to reconsideration of picture of synthesis of more heavy nuclei at energies of pycnonuclear reactions in compact stars.
So, the quasibound states are important
in understanding of synthesis of nuclei.


In Refs.~\cite{Maydanyuk_Shaulskyi.2022.EPJA,Maydanyuk.2023.Universe} 
interactions between nuclei are described on the basis of potential of interactions of Woods-Saxon type
used before in study of scattering of middle and heavy nuclei at low energies,
in study of bremsstrahlung in these reactions or fission.
There is attractive idea to implement description of interactions between two nuclei 
on the basis of fully many nucleon formalism.
Such an approach has been developed in nuclear physics for very long time~\cite{1937PhRv...52.1083W,1937PhRv...52.1107W}. 
We have ready formalism 
(see Refs.~\cite{Maydanyuk.2015.NPA,Maydanyuk_Zhang_Zou.2017.PRC} 
also \cite{Maydanyuk_Vasilevsky.2023.PRC,Shaulskyi_Maydanyuk_Vasilevsky.2024.PRC}, references therein)
to adapt it for the studied problem with interaction between nuclei from close distances 
(that takes place in nuclear processes in dense medium of compact stars).
Important advance of this approach is that interactions are calculated on the basis of two-nucleon interactions which have been studied deeply well.
Such a way allows to calculate all parameters of interactions between two nuclei,
not involving analysis of experimental cross sections of nuclear scattering to fix unknown parameters.
In contrary, parameters of potentials of Woods Saxon type are extracted from analysis of experimental cross sections of scattering of nuclei,
and are different for nuclei of different masses and energies.
In order to apply that for conditions of dense matter (where we have no experimental information from stars), 
this advance of many nucleon approach becomes important.
So, application of that formalism to study pycnonuclear reactions in compact stars is main idea of this paper.
Note that such approach has never been used in study of nuclear processes in compact stars.

In this paper, we are going to study interaction of two nuclei \isotope[12]{C} by using the folding
approximation of the cluster formalism. 
To achieve this aim, we have to select a proper wave
function, describing internal structure of \isotope[12]{C}, and an appropriate
 nucleon-nucleon interaction, which will determine dynamics of the
$\isotope[12]{C} + \isotope[12]{C}$ system.
%
Interest in studying of the reactions, induced by the $\isotope[12]{C} + \isotope[12]{C}$ interaction, is explained by its large
impact on the nucleosynthesis, the energy production and other important problems of stellar evolution~\cite{Gasques.2005.PRC,Chien.2018.PRC}.
These reactions havea significant impact on the evolution and structure of massive stars with mass larger than the solar mass. The $\isotope[12]{C} + \isotope[12]{C}$ fusion is known as pycnonulear reaction that reignites carbon-oxygen white dwarf into type Ia supernova explosion.
Such  reactions have been often studied to understand properties and evolution of white dwarfs.
Molecular structure of $^{24}$Mg considered as two-cluster system has long
history and a large number of methods have been used  
\cite{1975PhRvC..11.1803B, 1978PThPh..60.1013T,
1978PThPh..59..465K, 1978PThPh..59.1393A,
1980PThPS..68..303A, 1982PThPh..67..207O,
1982NuPhA.388..102S, 1989PThPh..81..390K,
1990NuPhA.513...75B, 1997PhRvC..55.1928K,
1998NuPhA.637..175K, 2018PhRvC..98f4604C} to study this structure.

\section{Choise of nuclei for analysis
\label{sec.analysis}}


\vspace{-1.5mm}
Hamada and Salpeter pointed out pycnonuclear reactions which in $10^{5}$ years transform some nuclei into others \cite{Hamada_Salpeter.1961.AstrJ}
(see also Ref.~\cite{ShapiroTeukolsky.2004.book}, p.~83, Sect.~3.5):
%
  (1) transformation of H to \isotope[4]{He} at density higher than $5 \cdot 10^{4}$ $\mbox{g} \cdot \mbox{cm}^{-3}$,
%
  (2) transformation of \isotope[4]{He} to \isotope[12]{C} at density higher than $8 \cdot 10^{8}$ $\mbox{g} \cdot \mbox{cm}^{-3}$,
%
  (3) transformation of \isotope[12]{C} to \isotope[24]{Mg} at density higher than $6 \cdot 10^{9}$ $\mbox{g} \cdot \mbox{cm}^{-3}$.
As Hamada and Salpeter estimated, \isotope[12]{C} are converted to \isotope[24]{Mg} in pycnonuclear reactions above a density of $6 \cdot 10^{9}$ ${\rm g} \cdot {\rm cm}^{-3}$
\cite{Hamada_Salpeter.1961.AstrJ}.
That was based on rates estimated
by Cameron~\cite{Cameron.1959b.AstrJ} 
(see improved calculations by Salpeter and Van Horn~\cite{Salpeter_VanHorn.1969.AstrJ}, etc.).
The critical density for H is of the order of $10^{6}$ $\mbox{g} \cdot \mbox{cm}^{-3}$,
for Carbon $5 \cdot 10^{10}$ $\mbox{g} \cdot \mbox{cm}^{-3}$.
Densities quoted here are still quite uncertain.
Finite temperatures and crystal imperfections increase the rates significantly
(the critical density for carbon can be about $5 \cdot 10^{10}$ ${\rm g} \cdot {\rm cm}^{-3}$).
But, difference in such estimations is not principal for analysis, so we will use the density  given by Hamada and Salpeter for calculations.



Let's denote distance between two nuclei located in neighboring lattice sites in the stellar medium as $2\, R_{0}$.
We define the density $\rho_{0}$ of matter in a sphere surrounding one nucleus of the lattice with radius $R_{0}$ as
the ratio of the mass of the nucleus to the volume of this sphere
($m_{u}$ is mass of the nucleon) as
$R_{0} = \bigl(A\, m_{u} \bigr)^{1/3} / \bigl( 4/3\, \pi\, \rho_{0} \bigr)^{1/3}$.
%
%
%
Here,
$R_{0}$ is the radius of the sphere that occupies one nucleus in lattice at the average density of the stellar medium $\rho_{0}$
($2 R_{0}$ is distance between two close nuclei located at the lattice sites).

We will place a new ``test'' nucleus between two neighboring nuclei in the lattice sites, considering it as a ``incident nucleus'' on one nucleus of the lattice.
The interaction potential between neighboring nuclei in the lattice sites is an external field in which this ``test'' nucleus is incident~\cite{ShapiroTeukolsky.2004.book}.
For $\isotope[12]{C} + \isotope[12]{C}$ we choose density as 
$\rho_{0} = 6 \cdot 10^{9}\, \displaystyle\frac{\mbox{g}} {\mbox{\rm cm}^{3}}$~\cite{ShapiroTeukolsky.2004.book}.
We find
$R_{0} = 92.5\; \mbox{\rm fm}$
and calculate concentration of nuclei as
$n_{A} = \rho_{0} / A\, m_{u}$
%
%
and obtain 
$n_{A} = 
3.\: 014 \cdot 10^{-7}\; 
\mbox{\rm fm}^{-3}$.

%




The interaction potential between two nuclei in Woods-Saxon form is defined as
%
%
$V (r) = v_{c} (r) + v_{N} (r) + v_{l=0} (r)$,
%
where $v_{c} (r)$, $v_{N} (r)$ and $v_{l} (r)$ are Coulomb, nuclear and centrifugal components
have the following form
%
\begin{equation}
\begin{array}{lll}
  \vspace{1mm}
  v_{N} (r) =  \displaystyle\frac{V_{R}} {1 + \exp{\displaystyle\frac{r-R_{R}} {a_{R}}}},
  \hspace{2mm}
  v_{l} (r) = \displaystyle\frac{l\,(l+1)} {2mr^{2}}, 
\end{array}
\label{eq.potentialWS.2a}
\end{equation}
%
%
\begin{equation}
\begin{array}{lll}
\vspace{1mm}
  v_{c} (r) =
  \left\{
  \begin{array}{ll}
    \displaystyle\frac{Z_{1} Z_{2} e^{2}} {r}, &
      \mbox{at  } r \ge R_{c}, \\
    \displaystyle\frac{Z_{1} Z_{2} e^{2}} {2 R_{c}}\;
      \biggl\{ 3 -  \displaystyle\frac{r^{2}}{R_{c}^{2}} \biggr\}, &
      \mbox{at  } r < R_{c}.
  \end{array}
  \right.
\end{array}
\label{eq.potentialWS.2}
\end{equation}
Here,
$m = m_{\rm p}\, A_{1} A_{2} / (A_{1} + A_{2})$ is reduced mass of two nuclei,
$m_{\rm p}$ is mass of nucleon (we use mass of proton in calculations),
$V_{R} = -75$ MeV is strength of nuclear component, and
%
%
%
\begin{equation}
\begin{array}{llll}
\vspace{1.0mm}
  R_{R} = r_{R}\, (A_{1}^{1/3} + A_{2}^{1/3}), &
  a_{R} = 0.44\; {\rm fm}, \\
  R_{c} = r_{c}\, (A_{1}^{1/3} + A_{2}^{1/3}), &
  r_{R} = r_c = 1.30\;{\rm fm}, 
\end{array}
\label{eq.potentialWS.4}
\end{equation}
where
$R_{c}$ and $R_{R}$ are Coulomb and nuclear radiuses of di-nuclear system, $a_{R}$ is diffusion parameter.
\begin{figure}[htbp]
\centerline{\includegraphics[width=88mm]{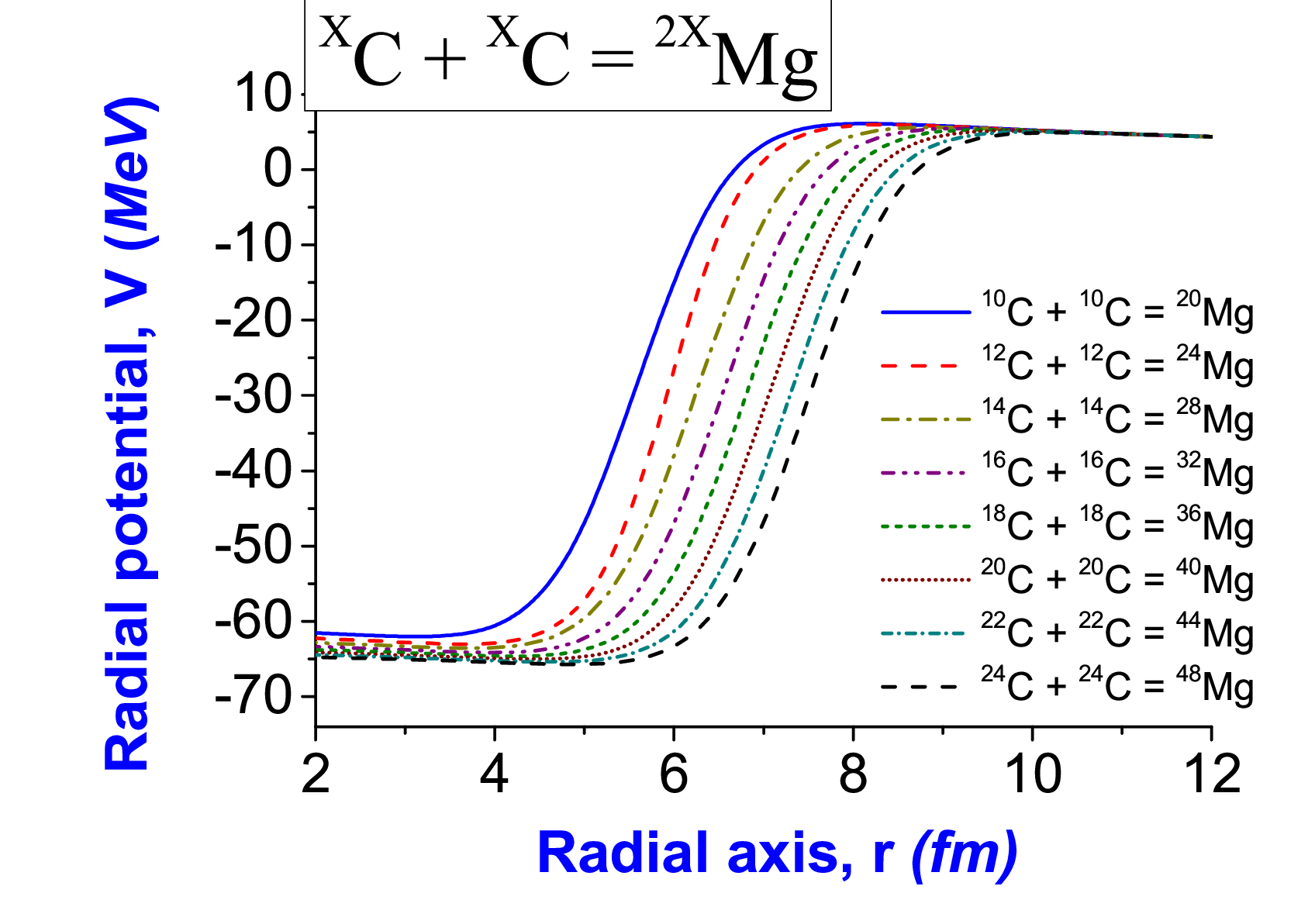}}
\vspace{-3mm}
\caption{\small (Color online)
Potential of interaction between two nuclei \isotope[12]{C}
[potential is defined in Eqs.~(\ref{eq.potentialWS.2a})--(\ref{eq.potentialWS.2})].
\label{fig.potentialWS.2}}
\end{figure}

Potentials for these reactions are shown in Fig.~\ref{fig.potentialWS.2}.
Barrier maxima and minima of the wells of the interaction potentials between isotope of carbon are shown in Tabl.~\ref{table.potentialWS.1}.
%
%
\begin{table}[htbp]
\begin{center}
\begin{tabular}{|c|c|c|c|c|c|c|c|c|c|c|} \hline
  Reaction &
    $r_{\rm min}$ & $V_{\rm min}$ & $r_{\rm max}$ & $V_{\rm max}$ & $R_{0}$ & $n_{A}$  \\ 
     &
     (fm) &  (MeV) & (fm) & (MeV) &  (fm) & ($10^{-7}\; {\rm fm}^{-3}$) \\
    \hline

  $\isotope[10]{C} + \isotope[10]{C}$ &
    3.36 & $-62.157$ & 7.98 & $6.249$ & 87.06 & 3.617\, 027\, 31 \\

  $\isotope[12]{C} + \isotope[12]{C}$ &
    3.64 & $-63.018$ & 8.33 & $5.972$ & 92.52 & 3.014\, 189\, 41 \\

  $\isotope[14]{C} + \isotope[14]{C}$ &
    3.92 & $-63.702$ & 8.68 & $5.743$ & 97.40 & 2.583\, 590\, 92 \\

  $\isotope[16]{C} + \isotope[16]{C}$ &
    4.20 & $-64.258$ & 8.96 & $5.552$ & 101.83 & 2.260\, 642\, 06 \\

  $\isotope[18]{C} + \isotope[18]{C}$ &
    4.48 & $-64.726$ & 9.24 & $5.386$ & 105.91 & 2.009\, 459\, 61 \\

  $\isotope[20]{C} + \isotope[20]{C}$ &
    4.62 & $-65.133$ & 9.52 & $5.242$ & 109.69 & 1.808\, 513\, 65 \\

  $\isotope[22]{C} + \isotope[22]{C}$ &
    4.90 & $-65.483$ & 9.80 & $5.115$ & 113.23 & 1.644\, 103\, 31 \\

  $\isotope[24]{C} + \isotope[24]{C}$ &
    5.04 & $-65.792$ & 10.08 & $5.001$ & 116.57 & 1.507\, 094\, 70 \\ \hline
\end{tabular}
\end{center}
\vspace{-1.5mm}
\caption{%
Parameters of minimum of the internal well and maximum of the barrier of the potential of interaction between Carbon isotopes,
the distance $R_{0}$ between the nuclei and concentration $n_{A}$
[such isotopes of Carbon were chosen in accordance with Ref.~\cite{Afanasjev.2010.ADNDT}
on the systematic study of astrophysical $S$-factors in fusion reactions of
\isotope{C}, \isotope{O}, \isotope{Ne}, \isotope{Mg},
the parameters were obtained for the density
$\rho_{0} = 6 \cdot 10^{9}\, \displaystyle\frac{\mbox{g}} {\mbox{\rm cm}^{3}}$].
}
\label{table.potentialWS.1}
\end{table}
%







\section{Formalism \label{Sec:MethodAM}}

The system of two \isotope[12]{C} nuclei is investigated in a two-cluster model. The
key element of this model is the a wave function describing internal structure
of each cluster. The main demands for such wave functions are following: 
(i) they should be antisymmetric with respect to permutation of a pair of
nucleons, 
(ii) they should be translationally invariant, and 
(iii) they should
properly (adequately) describe internal structure of each cluster.

We will employ many-particle shell model with oscillator potential to
construct wave function of internal motion of twelve nucleons. The lowest
shell-model wave function can be characterized by the Elliott quantum numbers
\cite{1958RSPSA.245..128E}, \cite{1958RSPSA.245..562E} $(\lambda\mu) = (04)$.
This quantum numbers indicates that \isotope[12]{C} has an oblate shape and thus it
has rotational band, which includes the 0$^{+}$ ground state and two excited
states $2^{+}$ and $4^{+}$.


\subsection{Channel classification.}

To describe system of two \isotope[12]{C} nuclei, we need to use six angular
variables and one radial variable. Two angular variables determine orientation
in the space of each \isotope[12]{C} nucleus, and two angular variables determine
relative orientation of two \isotope[12]{C} nuclei. As the result, we have to use six
quantum numbers related to rotations of each nucleus and two-cluster system as
well. Thus, we introduce two partial orbital momenta $l_{1}$ and $l_{2}$, they
are orbital momenta of the first and the second clusters. Then, we introduce
orbital momentum $l$ of relative rotation of both \isotope[12]{C} around their common
center of mass. Two additional quantum numbers are the total orbital momentum
$L$ and its projection $M$. Now we have to define the last sixth quantum
number. An orbital momentum $l_{0}$, which is a vector sum of the partial
angular momenta $\overrightarrow{l}_{0}=\overrightarrow{l}_{1}+\overrightarrow
{l}_{2}$ can be selected as this quantum number. With such definition, 
total orbital momentum $L$ is a vector sum of orbital momenta $l_{0}$ and $l$:
$\overrightarrow{L}=\overrightarrow{l}_{0}+\overrightarrow{l}$. 
Only radial variable is distance between two clusters \isotope[12]{C}.

Taking into account definition of these quantum numbers, wave function of two
interacting 
nuclei is 
%
\begin{equation}
\begin{array}{lcl}
\vspace{1.0mm}
  \Psi_{EL} & = & 
    \sum_{l,l_{1},l_{2},l_{0}}
    \Bigl[ 
      \left[  \Phi_{1} \left(^{12}\text{C},l_{1}\right)  \Phi_{2}\left(  ^{12}\text{C},l_{2}\right) \right]_{l_{0}} \times \\

  & &
  \psi_{EL;l}^{\left(  l_{1},l_{2},l_{0}\right)  } (r)  Y_{l} \left(  \widehat{\mathbf{r}}\right)  
  \Bigr] _{LM},
\end{array}
\label{eq:F001}
\end{equation}
where $\Phi_{1}$ and\ $\Phi_{2}$ are similar wave functions describing
internal structure of the first and second nuclei \isotope[12]{C}, they are
antisymmetric and translational invariant, wave function $\psi=\psi
_{EL;l}^{\left(  l_{1},l_{2},l_{c}\right)  }\left(  r\right)  $ describes
relative motion of clusters. As we neglect antisymmetrization in the compound
system, then the many-particle Schr\"{o}dinger equation is reduced to the set
of coupled equations for two-body system, with the\ Hamiltonian involved local
cluster-cluster potentials
\begin{equation}
\sum_{\widetilde{c}}\left[  \widehat{T}_{r}\delta_{c\widetilde{c}}+\widehat
{V}_{c\widetilde{c}}\left(  r\right)  \right]  \psi_{EL;\widetilde{c}}\left(
r\right)  =E\psi_{EL;c}\left(  r\right)  , \label{eq:F002}%
\end{equation}
with the multiple index $c=\left\{  l,l_{1},l_{2},l_{0} \right\} $ umbrageously numerates channels of two-cluster system. In general, quantum numbers $l,l_{1},l_{2},l_{0}$ are not the integrals of motion due to a nucleon-nucleon interaction. Such an approximation neglecting total antisymmetrization is called the folding approximation and local potential is called the direct or folding
potential. Definition of the folding potential and its explicit form will be
discussed in Sec.~\ref{Sec:FoldPot}. Now we proceed with the formulation of
our model. To solve numerically a system of Eq.~(\ref{eq:F002}), we
transform it to a system of linear algebraic equations
\begin{equation}
\sum_{\widetilde{c}}\sum_{m}\left[  \left\langle n\left\vert \widehat{T}%
_{r}\right\vert m\right\rangle \delta_{c\widetilde{c}}+\left\langle
n\left\vert \widehat{V}_{c\widetilde{c}}\right\vert m\right\rangle \right]
C_{mL;c}^{(E)}=EC_{nL;c}^{(E)}, \label{eq:F007}%
\end{equation}
by expanding wave functions $\psi_{EL;\widetilde{c}}\left(  r\right)  $ into a
complete basis of oscillator functions
\begin{equation}
\psi_{EL;c}\left(  r\right)  =\sum_{n}C_{nL;c}^{(E)}\Phi_{nL}\left(
r,b\right)  , \label{eq:F008}%
\end{equation}
where $b$ is the oscillator length (radius) and $\Phi_{nL}\left(  r,b\right)
$ is the oscillator wave function, explicit form of which can be found in Ref.~\cite{2015NuPhA.941..121L}. This method was formulated in
Refs.~\cite{kn:Fil_Okhr,kn:Fil81} and it is known as the algebraic
version of the resonating group method. In this method, the
expansion coefficients $\left\{  C_{nL;c}^{(E)}\right\}  $ are treated as wave
functions  of relative motion of clusters in oscillator representation, they
also treated as the wave function of the compound system since it can be
represented as
\begin{equation}
\begin{array}{lll}
\vspace{0.5mm}
  \Psi_{EL} & = &
  \sum_{c}\sum_{n}C_{nL;c}^{(E)}
    \Bigl[ 
      \left[ \Phi_{1}\left(^{12}\text{C},l_{1}\right)  
        \Phi_{2}\left(  ^{12}\text{C},l_{2}\right)
      \right]_{l_{0}} \\
  
  & \times &
  \Phi_{nL}\left(  r,b\right)  Y_{l} \left(  \widehat {\mathbf{r}}\right) 
  \Bigr]_{LM}, 
\end{array}
\label{eq:F009}%
\end{equation}
by taking into account Eq. (\ref{eq:F008}).


\subsection{Folding potential \label{Sec:FoldPot}}

In the folding approximation,
the inter-cluster potential is local and may be easily calculated, especially when simple
shell-model functions $\Phi _{i}(A_{i})$ are used to describe internal state
of clusters.
The folding potential is 
%
\begin{eqnarray}
  \widehat{V}^{(F)}(\mathbf{r})  &=& 
  \displaystyle\sum_{i\in A_{1}} \displaystyle\sum_{j\in A_{2}}
  \displaystyle\int dV_{1}\, dV_{2}\; \nonumber \\
  && \times |\Phi (A_{1})|^{2}\,
  \widehat{V}(\mathbf{r}_{i} - \mathbf{r}_{j})\: |\Phi (A_{2})|^{2},
\label{eq:F02}
\end{eqnarray}
where integration is performed over all coordinates
\begin{equation}
\begin{array}{llllll}
  dV_{1} = \displaystyle\prod\limits_{i\in A_{1}} d\mathbf{r}_{i}, &
  dV_{2} = \displaystyle\prod\limits_{i\in A_{2}} d\mathbf{r}_{i}
\end{array}
\end{equation}
and $\Phi (A_{\alpha })$ is a many-particle shell model function, describing
internal motion of $A_{\alpha }$ nucleons. 
As wave functions $\Phi _{1}\left( A_{1}\right) $ and$\ \Phi _{2}\left(
A_{2}\right) $ are translationally invariant, they actually depends on
coordinates
\begin{eqnarray}
\mathbf{r}_{i}^{\prime } &=& \mathbf{r}_{i}-\mathbf{R}_{1},\quad \mathbf{R}%
_{1}= \displaystyle\frac{1}{A_{1}} \displaystyle\sum_{i\in A_{1}}\mathbf{r}_{i},\quad i\in A_{1},
\label{eq:F03} \\
\mathbf{r}_{j}^{\prime } &=&\mathbf{r}_{j}-\mathbf{R}_{2},\quad \mathbf{R}%
_{2} = \displaystyle\frac{1}{A_{2}} \displaystyle\sum_{j\in A_{2}}\mathbf{r}_{j},\quad j\in A_{2},
\nonumber
\end{eqnarray}
respectively. 
%
%
This potential of interaction of two clusters (nuclei) is transformed to the following
%
%
\begin{equation}
  \widehat{V}^{(F)}(\mathbf{r}) =
  \displaystyle\int d\mathbf{r}_{1}\, d\mathbf{r}_{2}\,
  \rho_{1} \left( \mathbf{r}_{1}\right) \widehat{V}(\mathbf{r}_{1}-\mathbf{r}_{2} + \mathbf{r})\,
  \rho_{2} \left( \mathbf{r}_{2}\right).
\label{eq:F020}
\end{equation}
%
%
%
where $\widehat{V}\left(  \mathbf{r}_{1}-\mathbf{r}_{2}\right)  $ is a
nucleon-nucleon interaction and
\begin{equation}
\rho_{\alpha}\left(  \mathbf{r}\right)  =\int dV_{\alpha}^{\prime}\Phi
_{\alpha}\left(  A_{\alpha}\right)  \sum_{i=1}^{A_{\alpha}}\delta\left(
\mathbf{r-r}_{i}^{\prime}\right)  \Phi_{\alpha}\left(  A_{\alpha}\right)
\label{eq:F021}%
\end{equation}
is a single-particle density distribution in the first ($\alpha$=1) and second
($\alpha$=2) clusters. By using the Fourier transformation for nucleon-nucleon
interaction%
\begin{align}
\widehat{V}\left(  \mathbf{r}\right)   &  =\left(  2\pi\right)  ^{-3/2}\int
d\mathbf{k}\exp\left\{  -i\mathbf{k}\mathbf{r}\right\}  \mathcal{V}\left(
\mathbf{k}\right) \label{eq:F022}\\
\mathcal{V}\left(  \mathbf{k}\right)   &  =\left(  2\pi\right)  ^{-3/2}\int
d\mathbf{r}\exp\left\{  -i\mathbf{k}\mathbf{r}\right\}  \widehat{V}\left(
\mathbf{r}\right) \nonumber
\end{align}
we can reduce Eq.~(\ref{eq:F020}) to the form \
\begin{eqnarray}
\widehat{V}^{\left(  F\right)  }\left(  \mathbf{R}\right)  &=& \left(
2\pi\right)  ^{-3/2}\int d\mathbf{k}\mathcal{V}\left(  \mathbf{k}\right)
\exp\left\{  i\left(  \mathbf{k}\mathbf{R}\right)  \right\}  \nonumber \\
&& \times F_{1}\left(
\mathbf{k}\right)  F_{2}\left(  \mathbf{k}\right)  \label{eq:F024}%
\end{eqnarray}
where $F_{\alpha}\left(  \mathbf{k}\right)  $ is form factor determined as%
\begin{equation}
F_{\alpha}\left(  \mathbf{k}\right)  =\int dV_{\alpha}^{\prime}\Phi_{\alpha
}\left(  A_{\alpha}\right)  \sum_{i=1}^{A_{\alpha}}\exp\left\{  i\mathbf{k}%
\mathbf{r}_{i}\right\}  \Phi_{\alpha}\left(  A_{\alpha}\right).
\label{eq:F023}%
\end{equation}

We will use nucleon-nucleon potentials which have Gaussian spatial form:
\[
\widehat{V}\left(  ij\right)  =V_{0}\,\exp\left\{  -\frac{\left(
\mathbf{r}_{i}-\mathbf{r}_{j}\right)  ^{2}}{a^{2}}\right\}  .
\]
And the Fourier transform of the Gaussian potential is given by
\begin{equation}
\begin{array}{lllll}
\vspace{1.0mm}
  \widehat{V}\left(  \mathbf{r}\right) & = &
  V_{0}\,\exp\left\{ - \displaystyle\frac{\mathbf{r}^{2}}{a^{2}}\right\}  = 
  V_{0}\left(  \displaystyle\frac{a}{2} \right)^{3}
  \displaystyle \int
  d\mathbf{k}  \\
  && \times  \exp\left\{  -i\left( \mathbf{k}\mathbf{r}\right)  \right\}
  \exp\left\{  -\frac{a^{2}}{4}\mathbf{k}^{2}\right\} .
\end{array}
\label{eq:F004}%
\end{equation}

As was pointed out above, we employ the nuclear many-particle shell-model to
construct wave functions of \isotope[12]{C}. These wave functions as are constructed
as a Slater determinant from a single-particle orbital of three-dimensional
harmonic oscillator. For twelve nucleons or six protons and six neutron, we
need three orbitals:
\begin{align}
\left\vert 0\right\rangle  &  =\left\vert \mathbf{r}_{i},0\right\rangle
=\mathcal{N}_{0}\exp\left\{  -\frac{1}{2}\mathbf{r}_{i}^{2}\right\}
,\label{eq:F005A}\\
\left\vert \mathbf{u}_{\alpha}\right\rangle  &  =\left\vert \mathbf{r}%
_{i},\mathbf{u}_{\alpha}\right\rangle =\mathcal{N}_{\alpha}\left(
\mathbf{u}_{\alpha}\mathbf{r}_{i}\right)  \exp\left\{  -\frac{1}{2}%
\mathbf{r}_{i}^{2}\right\}  , \label{eq:F005B}%
\end{align}
where $\alpha=1,2,3$, and $\mathbf{u}_{1}$, $\mathbf{u}_{2}$ and
$\mathbf{u}_{3}$ is a set of three orthonormal unit vectors%
\begin{equation}
\left(  \mathbf{u}_{\alpha}\mathbf{u}_{\beta}\right)  =\delta_{\alpha\beta
},\quad\left[  \mathbf{u}_{\alpha}\mathbf{u}_{\beta}\right]  =\varepsilon
_{\alpha\beta\gamma}\mathbf{u}_{\gamma}.\nonumber
\end{equation}

Note that the orbital Eq.~(\ref{eq:F005B}) consists of the Hermitian polynomial
$H_{1}\left(  \mathbf{x}\right)  $ as%
\[
H_{1}\left(  \mathbf{u}_{\alpha}\mathbf{r}_{i}\right)  =2\left(
\mathbf{u}_{\alpha}\mathbf{r}_{i}\right).
\]

We use orbitals $\left\vert 0\right\rangle $, $\left\vert \mathbf{u}%
_{1}\right\rangle $ and $\left\vert \mathbf{u}_{2}\right\rangle $ to construct
wave functions, and we denote them as $\Phi_{1}\left(  ^{12}\text{C}\right)
=\Phi_{1}\left(  \left\{  \mathbf{u}_{\alpha}\right\}  ,^{12}\text{C}\right)
$. First, we need to calculate overlap of such functions. We present all
details of calculations in Appendix A, here we show the final results
\[
\begin{array}{lllll}
\vspace{1.0mm}
  \left\langle \Phi_{1}\left(  \left\{  \mathbf{u}_{\alpha}\right\}
,^{12}\text{C}\right)  |\Phi_{1}\left(  \left\{  \mathbf{u}_{\alpha}\right\}
,^{12}\text{C}\right)  \right\rangle \\
  = \left(  \left[  \mathbf{u}_{1}%
\mathbf{u}_{2}\right]  \left[  \widetilde{\mathbf{u}}_{1}\widetilde
{\mathbf{u}}_{2}\right]  \right)  ^{4}=\left(  \mathbf{u}_{3}\widetilde
{\mathbf{u}}_{3}\right)  ^{4},
\end{array}
\]
where $\widetilde{\mathbf{u}}_{1}$, $\widetilde{\mathbf{u}}_{2}$ and
$\widetilde{\mathbf{u}}_{3}$ is another set of unit vector with different
orientation with respect to vectors $\mathbf{u}_{1}$, $\mathbf{u}_{2}$ and
$\mathbf{u}_{3}$. Both sets of vectors will be used to project wave function
$\Phi_{1}\left(  \left\{  \mathbf{u}_{\alpha}\right\}  ,^{12}\text{C}\right)
$ onto the state with definite value of the orbital momentum $l_{1}$.

To calculate potential of the \isotope[12]{C}+\isotope[12]{C} interaction, we need to know
four form factors $F_{p\uparrow}\left(  \mathbf{k}\right)  $, $F_{p\downarrow
}\left(  \mathbf{k}\right)  $, $F_{n\uparrow}\left(  \mathbf{k}\right)  $,
$F_{n\downarrow}\left(  \mathbf{k}\right)  $, where an arrow indicates
projection of nucleon spin. As for \isotope[12]{C}, protons and neutrons with both
projections of spin occupy the same orbitals, then
\begin{eqnarray}
  F_{p\uparrow}\left(  \mathbf{k}\right) & = & 
  F_{p\downarrow}\left( \mathbf{k}\right) = 
  F_{n\uparrow}\left(  \mathbf{k}\right) = 
  F_{n\downarrow}\left(  \mathbf{k}\right) \nonumber \\  
  & = & 
  \displaystyle\frac{1}{3}\exp\left\{  -\displaystyle\frac{\left( kb\right)  ^{2}}{4}\,\frac{11}%
  {12}\right\}  
  \Bigl[ \left( 3 - \displaystyle\frac{1}{2}b^{2}\mathbf{k}^{2}\right)  \left(
  \mathbf{u}\widetilde{\mathbf{u}}\right) ^{4} + \nonumber \\
  & + &
  \frac{1}{2}b^{2}\left( \mathbf{uk}\right)  
  \left(  \mathbf{k}\widetilde{\mathbf{u}}\right)  \left(
  \mathbf{u}\widetilde{\mathbf{u}}\right)^{3} 
  \Bigr].  \label{eq:F010} 
\end{eqnarray}

By substituting the form-factor of \isotope[12]{C} (\ref{eq:F010}) \ into the
expression Eq.~(\ref{eq:F024}) and by using the Fourier transform Eq.~(\ref{eq:F004}), 
we obtain coordinate part of folding potential in the integral form
\begin{align}
&  \widehat{V}^{\left(  F\right)  }\left(  \mathbf{r}\right)  =\left(
\pi\right)  ^{-3/2}\int d\mathbf{k}\mathcal{V}\left(  \mathbf{k}\right)
\exp\left\{  i\left(  \mathbf{kr}\right)  \right\}  F_{1}\left(
\mathbf{k}\right)  F_{2}\left(  \mathbf{k}\right)  \nonumber \\
&  =\frac{1}{9}V_{0}\left(  \frac{a}{2}\right)  ^{3}\left(  \pi\right)
^{-3/2}\int d\mathbf{k}\exp\left\{  -\frac{a^{2}}{4}\gamma\mathbf{k}%
^{2}\right\}  \exp\left\{  i\left(  \mathbf{kr}\right)  \right\}  \nonumber\\
&  \times\left[  \left(  9-3b^{2}\mathbf{k}^{2}+\frac{1}{4}b^{4}\mathbf{k}%
^{4}\right)  \left(  \mathbf{u}\widetilde{\mathbf{u}}\right)  \left(
\mathbf{v}\widetilde{\mathbf{v}}\right)  ^{4}\right.  \nonumber\\
&  +\frac{1}{2}b^{2}\left(  3-\frac{1}{2}b^{2}\mathbf{k}^{2}\right)  \left(
\mathbf{uk}\right)  \left(  \mathbf{k}\widetilde{\mathbf{u}}\right)  \left(
\mathbf{u}\widetilde{\mathbf{u}}\right)  ^{3}\left(  \mathbf{v}\widetilde
{\mathbf{v}}\right)  \nonumber\\
&  +\frac{1}{2}b^{2}\left(  3-\frac{1}{2}b^{2}\mathbf{k}^{2}\right)  \left(
\mathbf{vk}\right)  \left(  \mathbf{k}\widetilde{\mathbf{v}}\right)  \left(
\mathbf{u}\widetilde{\mathbf{u}}\right)  ^{4}\left(  \mathbf{v}\widetilde
{\mathbf{v}}\right)  ^{3} \nonumber\\
&  \left.  +\frac{1}{4}b^{4}\left(  \mathbf{uk}\right)  \left(  \mathbf{k}%
\widetilde{\mathbf{u}}\right)  \left(  \mathbf{vk}\right)  \left(
\mathbf{k}\widetilde{\mathbf{v}}\right)  \left(  \mathbf{u}\widetilde
{\mathbf{u}}\right)  ^{3}\left(  \mathbf{v}\widetilde{\mathbf{v}}\right)
^{3}\right]  , \label{eq:F012}
\end{align}
with $\gamma=1+\frac{11}{6}\frac{b^{2}}{a^{2}}$ and $\mathbf{v}$ ($\widetilde{\mathbf{v}}$)\ is a unit vector determining the
orientation of the second \isotope[12]{C} cluster. By performing tedious but
straightforward calculations of the integral over vector $k$, we obtain
expression for the folding potential of interaction of two \isotope[12]{C}.
Obviously, this potential has a tensor form, as it depends on mutual
orientation of clusters. It can be symbolically presented as
\begin{equation}
\begin{array}{lll}
\vspace{1.5mm}
  \widehat{V}^{\left(  F\right)  }\left(  \mathbf{r}\right) & = &
  \displaystyle\sum_{\lambda=0,2,4} \displaystyle\sum_{\mu=-\lambda}^{\lambda}
    \widehat{V}_{\lambda}^{\left( F\right)  }\left( r\right)  Y_{\lambda\mu}\left( \widehat{\mathbf{r}} \right) \\  

  & \times &
  \Bigl\{  
    \mathcal{Y}_{c}\left(  \left\{  \mathbf{u,v}\right\} \right)
    \mathcal{Y}_{\widehat{c}}\left(  \left\{  \widetilde{\mathbf{u}} \mathbf{,} \widetilde{\mathbf{v}}\right\} \right) 
  \Bigr\}_{\lambda\mu},
\end{array}
\label{eq:F015}
\end{equation}
where is a spherical harmonics, $\mathcal{Y}_{c}\left(  \left\{
\mathbf{u,v}\right\}  \right) $ is an angular part of the wave function
(\ref{eq:F009}) and $c$ is the index (\ref{eq:F003}) of the \isotope[12]{C}+\isotope[12]{C}
channel. The folding potential (\ref{eq:F015}) contains the scalar component
$\lambda=0$ and two tensor components with $\lambda=2$ and $\lambda=4$. \ In
the present paper, we restrict ourselves to the scalar part of the folding
potential. In this case, partial angular momenta $l_{1}$ and $l_{2}$ are
integral of motion. Moreover, in this case the quantum number $l_{0}$ is
redundant and the total orbital momentum $L$ is equal to the orbital momentum
$l$ of relative motion of two clusters.

By using results obtained for the folding potential generated by
nucleon-nucleon interaction, we can easily construct the folding potential
generated the Coulomb interaction of protons. For this aim we use the integral
which relates the Coulomb interaction and Gaussian potential
\begin{equation}
\frac{1}{r}=\frac{2}{\sqrt{\pi}}\int_{0}^{\infty}dx\exp\left\{ -r^{2}%
x^{2}\right\}  .\label{eq:F017}%
\end{equation}

As the results for the Coulomb interaction, we obtain formula which are
similar to formula Eq.~(\ref{eq:F012}) with the additional integration over the
range of Gaussian potential. We do not display this formula, as well as
results of integration over the vector $\mathbf{k}$ in Eq.~(\ref{eq:F012}). Here
we display only\ scalar part of the nuclear and Coulomb interactions of two
\isotope[12]{C} cluster.

Nucleon-nucleon part of the \isotope[12]{C}+\isotope[12]{C} interaction can be represented
as
%
\begin{equation}
\begin{array}{lllll}
\vspace{1.5mm}
  \widehat{V}_{NN}^{(F)}\left( \mathbf{r}\right) & = &
  9 \displaystyle\sum_{\nu=1}^{N_{G}}\left[  9V_{33}^{\left(  \nu\right)  }+3V_{31}^{\left(  \nu\right) }%
  +3V_{13}^{\left(  \nu\right)  } +V_{11}^{\left(  \nu\right)  }\right] \\

\vspace{1.5mm}
  & \times &
  z_{\nu}^{3/2}\exp\left\{ - \displaystyle\frac{\mathbf{r}^{2}}{a_{\nu}^{2}}z_{\nu}\right\} \\ 

\vspace{1.5mm}
  & \times &
  \left\{  \left[  1- \displaystyle\frac{4}{3}\left(  \displaystyle\frac{b}{a_{\nu}}\right)
  ^{2}z_{\nu} + \displaystyle\frac{20}{27}\left( \displaystyle\frac{b}{a_{\nu}}\right)^{4} z_{\nu}%
  ^{2}\right] \right. \\

\vspace{1.5mm}
  & + & 
  \left(  \displaystyle\frac{b}{a_{\nu}}\right)  ^{2}z_{\nu}^{2}\left[  \displaystyle\frac{8}{9}%
  - \displaystyle\frac{80}{81}\left( \displaystyle\frac{b}{a_{\nu}}\right)  ^{2}z_{\nu}\right]  \left(
  \displaystyle\frac{\mathbf{r}^{2}}{a_{\nu}^{2}}\right)  \\

  & + & 
  \left. \displaystyle\frac{16}{81}\left( \displaystyle\frac{b}{a_{\nu}}\right)  ^{4}z_{\nu}%
  ^{4}\left( \displaystyle\frac{\mathbf{r}^{4}}{a_{\nu}^{4}}\right)  \right\},
\end{array}
\label{eq:R021}
\end{equation}
%
where $\mathbf{r}$ is a distance between center of mass of two nuclei $^{12}%
$C, $a_{\nu}$ is a range (diffuseness) of NN potential, and $z_{\nu}=\left(  1+\frac{11}{6}\frac{b^{2}}{a_{\nu}^{2}}\right)  ^{-1}$. Expression Eq.~(\ref{eq:R021}) is deduced for a nucleon-nucleon potential which is
represented as
\begin{equation}
\begin{array}{lllll}
\vspace{1.5mm}
  V_{NN}^{\left(  C\right)  }\left(  \mathbf{r}_{i}-\mathbf{r}_{j}\right) & = &
  \displaystyle\sum_{S=0,1} \displaystyle\sum_{T=0,1} \displaystyle\sum_{\nu=1}^{N_{G}} V_{2S+1,2T+1}^{\left( k\right) } \\
  
  & \times &
  \exp\left\{ - \displaystyle\frac{\left(  \mathbf{r}_{i}-\mathbf{r}_{j}\right) ^{2} }{a_{\nu}^{2}}\right\} \widehat{P}_{S}\widehat{P}_{T}, 
\end{array}
\label{eq:R025}
\end{equation}
where $\widehat{P}_{S}$\ ($\widehat{P}_{T}$) is the projection operator
selecting the spin $S$ (isospin $T$) of interaction nucleons. One see that the
selected nucleon-nucleon potential is a superposition of several ($N_{G}$) Gaussians.

The Coulomb interaction of protons generates the corresponding part of the
folding potential for \isotope[12]{C}+\isotope[12]{C} system:
\begin{align}
&  \widehat{V}_{C}^{\left(  F\right)  }\left(  r\right)  =\frac{Z_{1}%
Z_{2}e^{2}}{r}\frac{1}{\sqrt{\pi}}\label{eq:R028}\\
&  \times\frac{1}{9}\left[  9\gamma\left(  \frac{1}{2},\frac{r^{2}}{\sigma
b^{2}}\right)  -12\left(  \frac{b}{r}\right)  ^{2}\gamma\left(  \frac{3}%
{2},\frac{r^{2}}{\sigma b^{2}}\right)  \right. \nonumber\\
&  +8\left(  \frac{b}{r}\right)  ^{2}\gamma\left(  \frac{5}{2},\frac{r^{2}%
}{\sigma b^{2}}\right)  +\frac{20}{3}\left(  \frac{b}{r}\right)  ^{4}%
\gamma\left(  \frac{5}{2},\frac{r^{2}}{\sigma b^{2}}\right) \nonumber\\
&  -\frac{80}{9}\left(  \frac{b}{r}\right)  ^{4}\gamma\left(  \frac{7}%
{2},\frac{r^{2}}{\sigma b^{2}}\right)  +\left.  \frac{16}{9}\left(  \frac
{b}{r}\right)  ^{4}\gamma\left(  \frac{9}{2},\frac{r^{2}}{\sigma b^{2}%
}\right)  \right]  ,\nonumber
\end{align}
where $\sigma=\frac{11}{6}$ and $\gamma\left(  a,z\right) =a^{-1}z^{a}e^{-z}M\left(  1,1+a;z\right)$ is the incomplete gamma function.


For the nucleon-nucleon part of folding potential, the deformed form of clusters stipulates that it has Gaussian form multiplied on
polynomial of second order of the variable $\mathbf{r}^{2}$, the square
distance between clusters. in the case of the Coulomb part of folding
potential, the deformation of cluster generates the superposition of the
incomplete gamma functions. If we restrict ourselves only with the first
terms in Eqs.~(\ref{eq:R021} ) and (\ref{eq:R028}), i.e. if we approximate
the nucleon-nucleon part of folding potential as
\begin{equation}
\begin{array}{llllll}
\vspace{1.0mm}
  \hat{V}_{NN}^{(F)}(\mathbf{r}) & = & 
  9\,\displaystyle\sum_{\nu =1}^{N_{G}}
  \Bigl(9V_{33}^{(\nu )}+3V_{31}^{(\nu )}+3V_{13}^{(\nu )}+V_{11}^{(\nu )} \Bigr) \\
  
  & \times & 
  z_{\nu }^{3/2}\exp \Bigl\{-\displaystyle\frac{\vb{r}^{2}}{a_{\nu}^{2}}\,z_{\nu }\Bigr\},
\end{array}
\label{eq.fold.nuclei-s.potential.C12}
\end{equation}
and the Coulomb part as
\begin{equation}
\widehat{V}_{C}^{\left( F\right) }(\mathbf{r})=\displaystyle\frac{%
Z_{1}\,Z_{2}\,e^{2}}{R}\;\mathrm{erf}\left( \displaystyle\frac{r^{2}}{\sigma
\,b^{2}}\right) ,  \label{eq:F13C}
\end{equation}%
where 
\begin{equation}
a=\displaystyle\frac{\sigma \gamma ^{2}}{1+\sigma \gamma ^{2}},\,\sigma
=2-\mu ^{-1},\mu =\frac{A_{1}A_{2}}{A_{1}+A_{2}}.  \label{eq:F11asigma}
\end{equation}

\begin{figure}[htbp]
\centerline{\includegraphics[width=88mm]{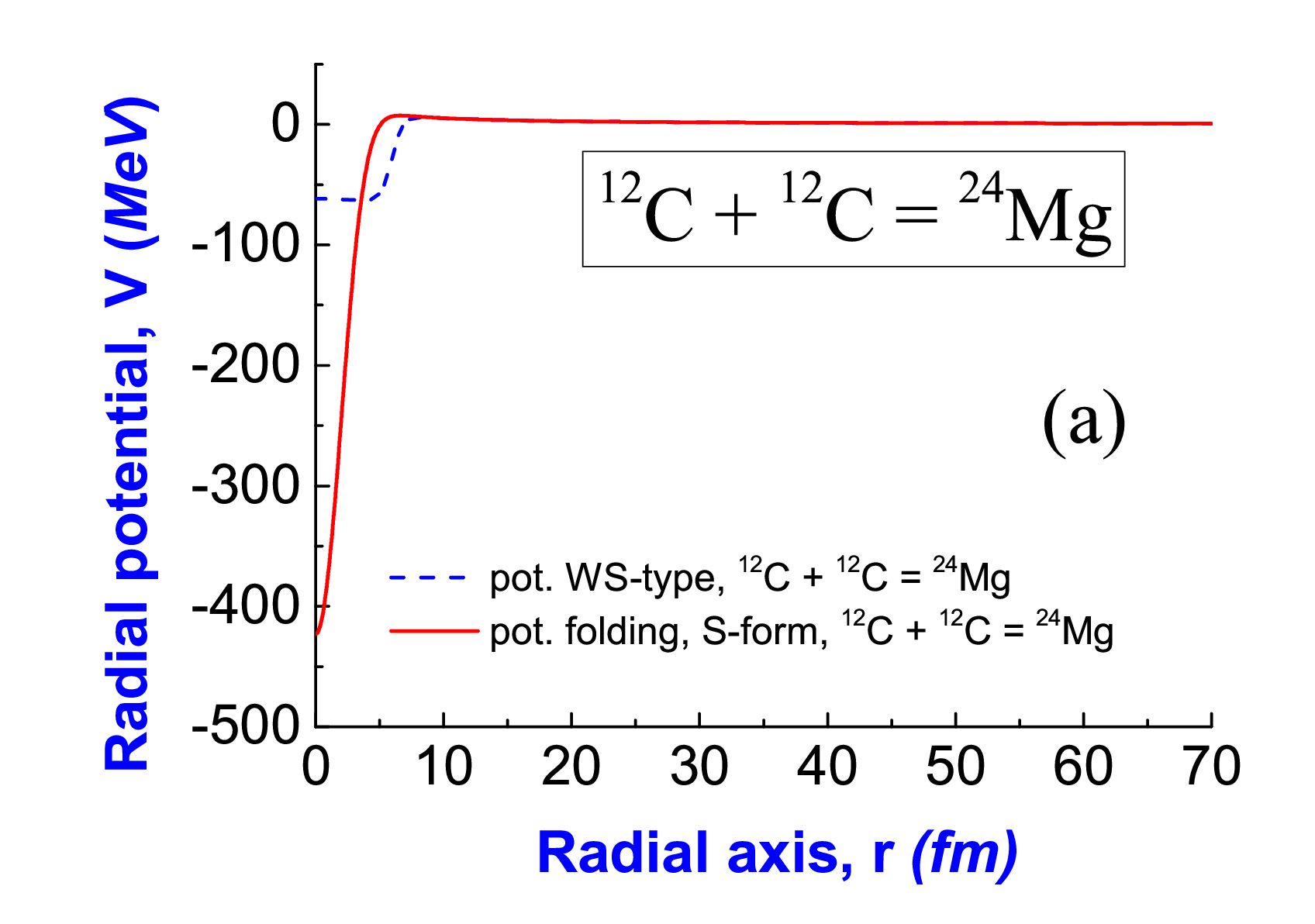}}
\centerline{\includegraphics[width=88mm]{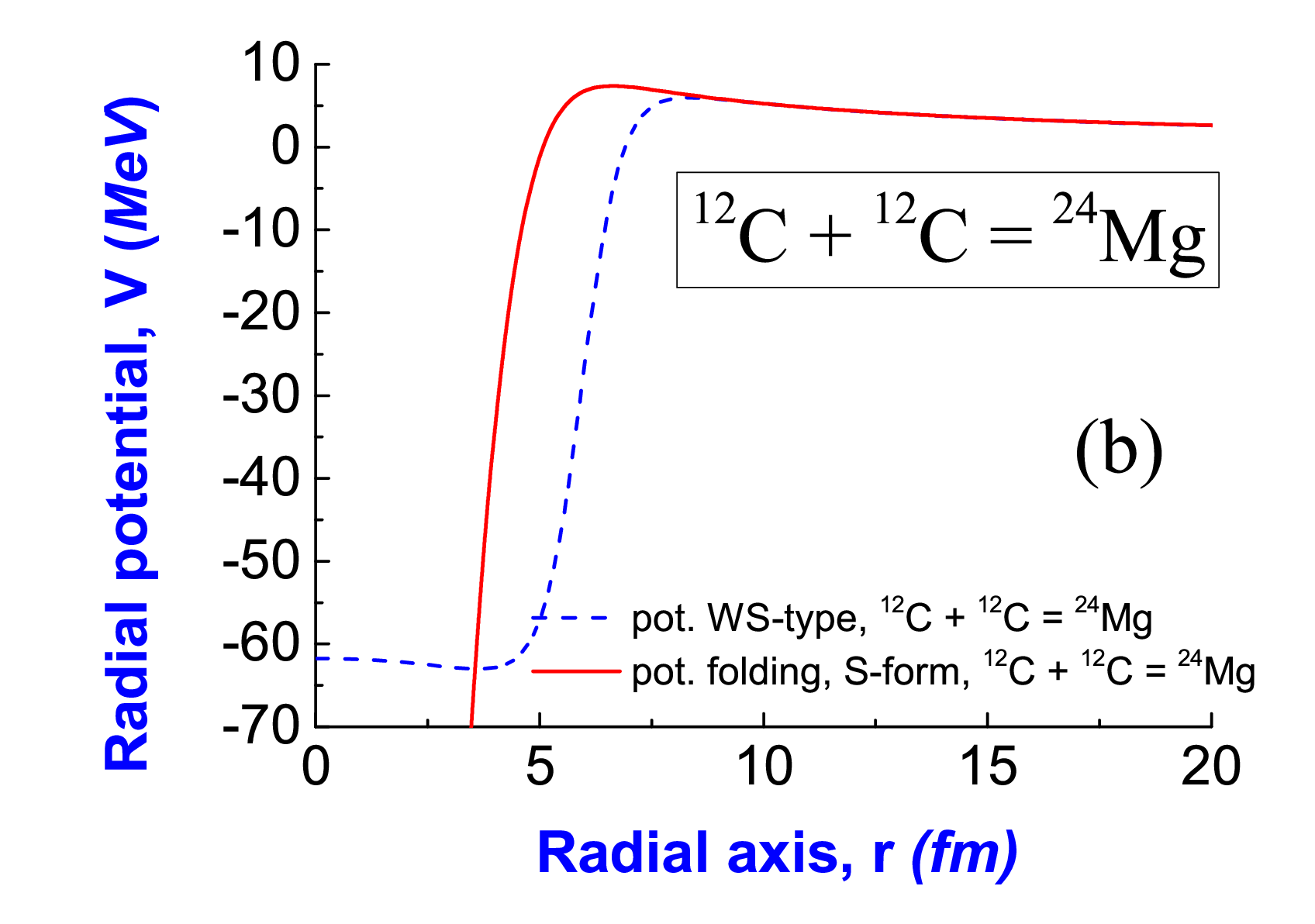}}
\vspace{-3mm}
\caption{\small (Color online)
The potential of Woods-Saxon type and the folding potential with the S-form. 
%
\label{fig.fold.potential.1}}
\end{figure}

Recall that the
deformation of $\isotope[12]{C}$ is formed by eight nucleons residing on
p-shell orbitals, while four nucleons occupying s-shell orbitals do not
deform the nucleus. Note that the simplified form of nucleon-nucleon
potential of Eq.~(\ref{eq.fold.nuclei-s.potential.C12}) within the simple factor
coincides with the full form of folding potential for interaction of the
s-shell clusters, and the simplified form of the Coulomb interaction of Eq.~(\ref%
{eq:F13C}) coincides with the full form for interaction of the s-shell
clusters.
%
%
The folding potential 
with S-form in comparison with the potential of Woods-Saxon
type are shown in Fig.~\ref{fig.fold.potential.1}.

%







\begin{figure}[htbp]
\centerline{\includegraphics[width=82mm]{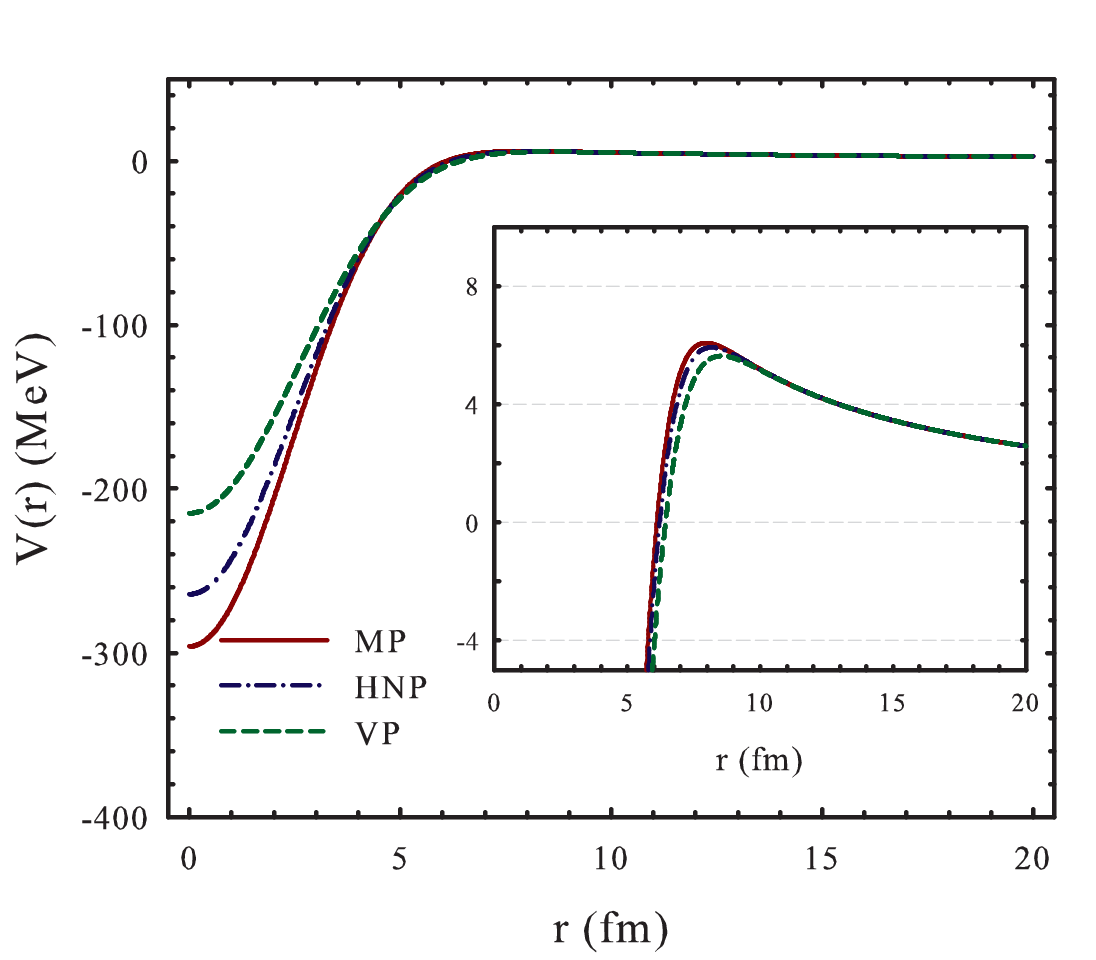}}
\vspace{-2mm}
\caption{\small
Folding potentials for the \isotope[12]{C}+\isotope[12]{C} system generated by the
Minnesota and Hasegawa-Nagata potentials.}
\label{Fig:FoldPoten12C+12C}
\end{figure}

In Fig.~\ref{Fig:FoldPoten12C+12C} we display folding potential of Eq.~(\ref{eq:R021}) 
for \isotope[12]{C}+\isotope[12]{C} system generated by Minnesota (MP)
\cite{kn:Minn_pot1}, Hasegawa-Nagata (HNP) \cite{potMHN1, potMHN2} and
Volkov (VP) \cite{kn:Volk65} potentials. These potentials are often used in
many-cluster systems and in quasi-molecular systems as well. In all cases we
use the same oscillator length $b = 1.783$~fm, it means the size (mass root mean
square radius) of \isotope[12]{C} is the same for these potentials. The all
potentials generates similar folding potentials which almost coincide at the
rather large range of inter-cluster distance. The noticeable difference of
folding potentials is observed only at small distances: $r \leq 3$~fm. In Fig.
\ref{Fig:FoldPoten12C+12C} we also show in detail a barrier which is created
by the Coulomb interactions. The shape and \ height of barrier is slightly
depend on the shape of nucleon-nucleon potential.%

%

%

\begin{figure}[htbp]
\centerline{\includegraphics[width=88mm]{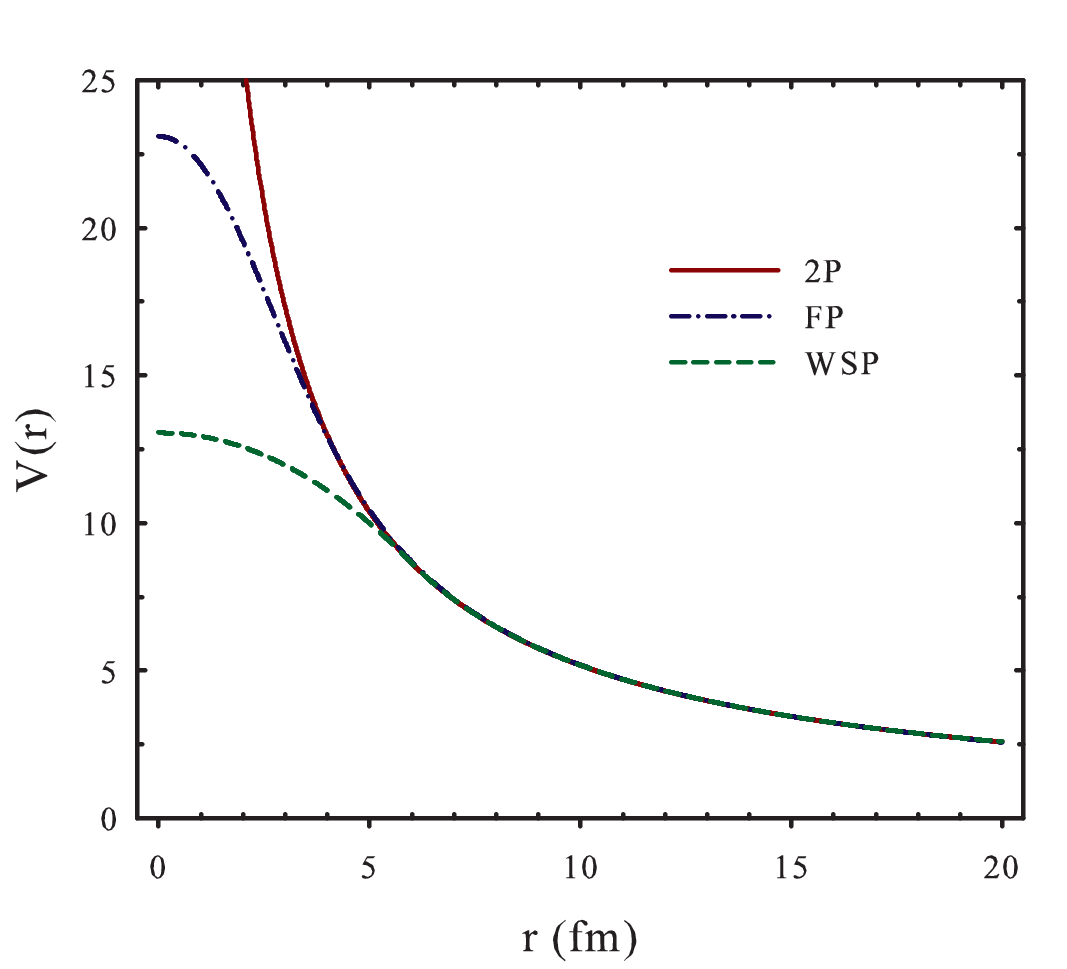}}
\vspace{-2mm}
\caption{\small
The part of inter-cluster potential which is related to the Coulomb interaction of \isotope[12]{C} clusters.}
\label{Fig:CoulPotens}
\end{figure}

Let us now consider the inter-cluster potential which is generated by the
Coulomb interaction. In Fig.~\ref{Fig:CoulPotens} we compare the Coulomb
inter-cluster potential, which is suggested the folding model (FP), with the
potential which is used with the Woods-Saxon potential (WSP). In Fig.
\ref{Fig:CoulPotens} we also display the Coulomb potential of two
structureless particles (2P) with the charges $Z_{1} = Z_{2} = 6$. One can see,
that the FP\ Coulomb potential is more stronger than the WSP\ Coulomb
potential when the distance between two \isotope[12]{C} clusters is less than 5 fm.
When this distance is larger 6 fm, these potentials are equal and coincide
with two-particle (2P) potential.

%

\section{Resonance states in \isotope[12]{C}+\isotope[12]{C} system
\label{sec.cluster.resonance}}

To solve system of Eq.~(\ref{eq:F002}) and to calculate spectrum of
resonance states in \isotope[24]{Mg}, we need to fix the oscillator length $b$, which stands for the
distribution of nucleons inside \isotope[12]{C}. We selected such value of $b$ which
reproduces the mass root-mean-square radius of \isotope[12]{C}. If we take $b = 1.738$~fm, 
then with the shell-mode wave functions we obtain the experimental value
\cite{KELLEY201771} $R_{m} = 2.4829$~fm of mass root-mean-square radius.

\begin{figure}[htbp]
\centerline{\includegraphics[width=88mm]{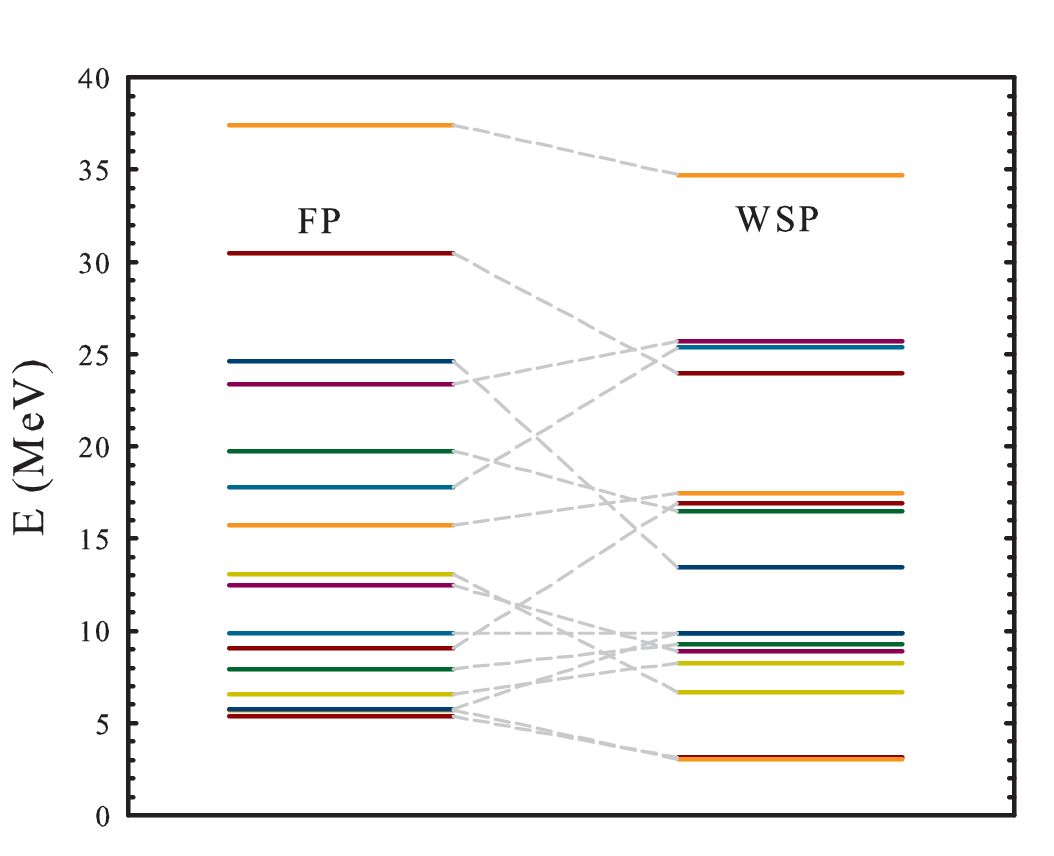}}
\caption{\small (Color online)
Position of the $^{24}$Mg resonance states.
\label{Fig:SpectrFPvsWSP}}
\end{figure}

In Fig.~\ref{Fig:SpectrFPvsWSP} we demonstrate spectrum of the resonance
states, obtained with the folding and Woods-Saxon potentials. One can see that
the folding potential generates almost uniformly distributed resonance states
over the selected energy range. From other side, the Woods-Saxon potential
created several groups of closely positioned\ resonance states: around 8, 16
and 24 MeV.

In Table \ref{Tab:ResonsVP} we collect parameters of resonance states obtained
with the Volkov potential \cite{kn:Volk65} and with the Woods-Saxon potential,
with parameters from Ref.~\cite{Maydanyuk.2023.Universe}. 
We display energies and widths of resonance states, and mass root-means square radii and average
distances between clusters in these states. One can see that two interacting
\isotope[12]{C} create fairly narrow resonance states. Starting from 10$^{+}$ state
to $18^{+}$ states, the folding potential creates two resonance states, one of
them is very narrow width the width less than 20 keV and another is fairly wide
width $\Gamma\approx$2.0 MeV. The Woods-Saxon potential also generates two
resonance states from $12^{+}$ to 20$^{+}$.%
%
\begin{table*}[htbp] \centering
\caption{Energies and widths of resonance states in $^{24}$Mg above the
\isotope[12]{C}+\isotope[12]{C} threshold.}%
\begin{tabular}
[c]{|c|c|c|c|c|c|c|c|c|}\hline
& \multicolumn{4}{|c}{FP} & \multicolumn{4}{|c|}{WSP}\\\hline
$J^{\pi}$ & $E$, MeV & $\Gamma$, MeV & $R_{m}$, fm & $A_{c}$, fm & $E$, MeV &
$\Gamma$, MeV & $R_{m}$, fm & $A_{c}$, fm\\\hline
0$^{+}$ & 5.360 & 0.267 & 5.967 & 4.579 & 3.139 & 0.0286 & 3.423 &
2.245\\\hline
2$^{+}$ & 5.720 & 0.331 & 6.342 & 4.904 & 3.040 & 0.1622 & 3.383 &
2.205\\\hline
4$^{+}$ & 6.573 & 0.493 & 6.610 & 5.134 & 8.267 & 13.0405 & 8.055 &
6.363\\\hline
6$^{+}$ & 7.948 & 0.771 & 6.988 & 5.458 & 9.256 & 3.3746 & 8.117 &
6.415\\\hline
8$^{+}$ & 9.890 & 1.160 & 7.019 & 5.484 & 9.861 & 0.9546 & 6.811 &
5.306\\\hline
10$^{+}$ & 5.749 & 0.0198 & 3.135 & 1.945 & 9.897 & 0.1053 & 5.775 &
4.412\\\hline
10$^{+}$ & 12.457 & 1.6187 & 7.080 & 5.536 & - & - &  & \\\hline
$12^{+}$ & 9.076 & 0.0105 & 3.107 & 1.915 & 8.893 & 0.0010 & 3.345 &
2.166\\\hline
$12^{+}$ & 15.717 & 2.0694 & 7.000 & 5.468 & 16.943 & 5.1233 & 7.861 &
6.199\\\hline
14$^{+}$ & 13.076 & 0.0013 & 3.087 & 1.894 & 17.446 & 0.8560 & 6.266 &
4.838\\\hline
14$^{+}$ & 19.746 & 2.4092 & 7.168 & 5.611 & 6.6730 & 0.0013 & 3.255 &
2.072\\\hline
16$^{+}$ & 17.815 & 0.0030 & 3.220 & 2.036 & 16.481 & 0.0184 & 3.603 &
2.426\\\hline
16$^{+}$ & 24.628 & 2.5313 & 7.076 & 5.533 & 25.367 & 7.0115 & 7.904 &
6.234\\\hline
18$^{+}$ & 23.371 & 0.0032 & 4.386 & 3.171 & 13.435 & 0.4703 & 3.319 &
2.139\\\hline
18$^{+}$ & 30.465 & 2.3253 & 6.890 & 5.374 & 25.682 & 1.0205 & 6.346 &
4.908\\\hline
20$^{+}$ & 37.418 & 1.6296 & 6.329 & 4.892 & 23.960 & 0.0117 & 3.692 &
2.514\\\hline
20$^{+}$ & - & - &  &  & 34.710 & 5.2244 & 7.641 & 6.012\\\hline
\end{tabular}
\label{Tab:ResonsVP}%
\end{table*}%


In Table \ref{Tab:ResonsVP} the mass root-means-square radius $R_{m}$
and the average distance between clusters $A_{c}$ are determined as
\begin{align}
R_{m}  &  =\sqrt{\left\langle \Psi_{EL}\left\vert \sum_{i=1}^{A}\mathbf{r}%
_{i}^{2}\right\vert \Psi_{EL}\right\rangle /A},\label{eq:R01A}\\
A_{c}  &  =\sqrt{\left\langle \psi_{EL}\left\vert \mathbf{r}^{2}\right\vert
\psi_{EL}\right\rangle }, \label{eq:R01B}%
\end{align}
where $\mathbf{r}_{i}$\ is the single-particle coordinate, $\mathbf{r}$ is the
distance between center of mass of clusters \isotope[12]{C}, $A$= 24 is the total
number of nucleons. Eqs. (\ref{eq:R01A}) and (\ref{eq:R01B}) involve the total
wave function $\Psi_{EL}$ of $^{24}$Mg considered as a two-cluster system, and
wave function of relative motion of clusters $\psi_{EL}=\psi_{EL}\left(
\mathbf{r}\right)  $, which is a function of the vector $\mathbf{r}$
determining relative position of clusters. Eqs. (\ref{eq:R01A}) and
(\ref{eq:R01B}) are usually employed to study bound states only, as for the
continuous spectrum states corresponding matrix elements are divergent.
However, an algorithm is suggested in Ref. \cite{2023UkrJPh..68..3K} to extend
these quantities for investigation of the continuous spectrum states and
resonance states especially. In oscillator representation, the quantities
$R_{m}$ and $A_{c}$ are determined in the following way%
\begin{align}
R_{m}  &  =\sqrt{\sum_{n,m=0}^{N_{0}}C_{n,L}^{E}\left\langle n,L\left\vert
\sum_{i=1}^{A}\mathbf{r}_{i}^{2}\right\vert m,L\right\rangle C_{m,L}^{E}%
/A},\label{eq:R02A}\\
A_{c}  &  =\sqrt{\sum_{n,m=0}^{N_{0}}C_{n,L}^{E}\left\langle n,L\left\vert
\mathbf{r}^{2}\right\vert m.L\right\rangle C_{m,L}^{E}}, \label{eq:R02B}%
\end{align}
where $N_{0}$ is the number of oscillator functions involved in calculations.
As was shown in Ref. \cite{2023UkrJPh..68..3K}, with such definitions of
$R_{m}$ and $A_{c}$, given in Eqs. (\ref{eq:R02A}) and (\ref{eq:R02B}),
\ these quantities can be calculated not only bound states, but also
continuous spectrum states, provided that the expansion coefficients $\left\{
C_{n,L}^{E}\right\}  $ are obtained by diagonalizing the $N_{0}\times N_{0}$
matrix of the Hamiltonian.

To understand peculiarities of cluster-cluster interaction and to obtain
additional information about resonance state, we introduce new parameter (also
use) the weight of the internal part of a wave function\ of continuous
spectrum state. It defined as%
\[
W_{int}\left(  E_{\alpha}\right)  =\sum_{n=0}^{N_{i}}\left\vert C_{n,L}%
^{E_{\alpha}}\right\vert ^{2},
\]
where $E_{\alpha}$ is the eigenenergy of the two-cluster Hamiltonian,
coefficients $\left\{  C_{n,L}^{E_{\alpha}}\right\}  $\ is the corresponding
eigenfunction of the Hamiltonian in the oscillator representation. Parameter
$N_{i}$\ determines the the total number of oscillator functions describing
the internal region. In our calculations, we selected $N_{i}=30$, which allows
us to cover the intercluster distances $0\leq r<20$ fm.

\begin{figure}[htbp]
\centerline{\includegraphics[width=88mm]{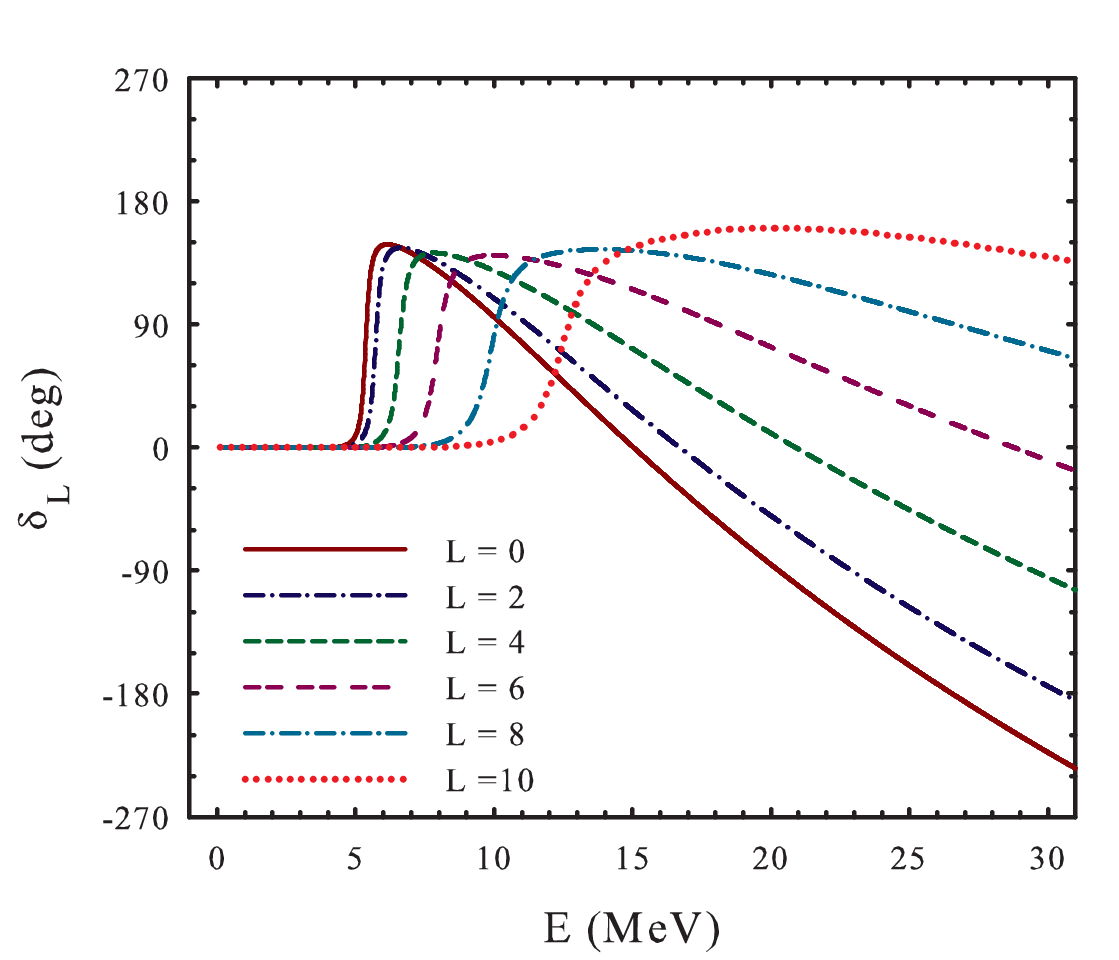}}
\caption{\small (Color online)
Phase shifts of the elastic scattering of \isotope[12]{C} on \isotope[12]{C} as a
function of energy. Phase shifts obtained in the folding approximation with the Volkov potential
\label{Fig:PhasesFPV}}
\end{figure}

Phase shifts of the elastic \isotope[12]{C}+\isotope[12]{C} scattering for $L$=0, 2, 4, 6,
8, and 10, which are shown in Fig. \ref{Fig:PhasesFPV}. The smaller is the width of a resonance
state, the faster is growing of the corresponding phase shifts. It is
important to note that phase shifts for all states displayed in Fig.
\ref{Fig:PhasesFPV} do not reach 180$^{\circ}$, as one should expect analyzing
the Breit-Wigner formula. Such behavior of the phase shifts indicates that the
so-called potential or background phase shifts rapidly decreasing with
increasing of energy. It can be seen in Fig.~\ref{Fig:PhasesFPV} that one observes 
the behavior of phase shifts far from resonance energy.%

%

\begin{figure}[htbp]
\centerline{\includegraphics[width=88mm]{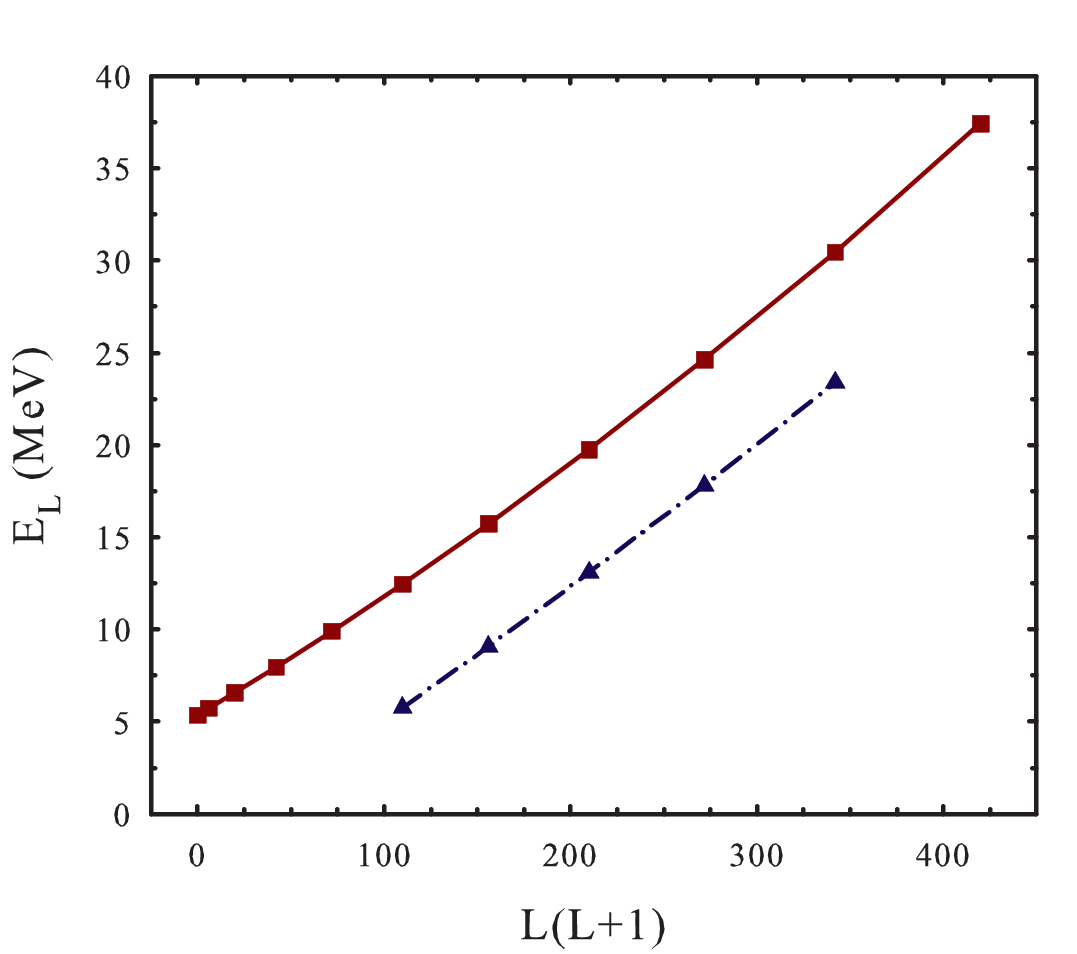}}
\caption{\small (Color online)
Energy of resonance states as a function of the factor $L\left(L+1\right)$.
\label{Fig:EnergyResonsvsL}}
\end{figure}
\begin{figure}[htbp]
\centerline{\includegraphics[width=88mm]{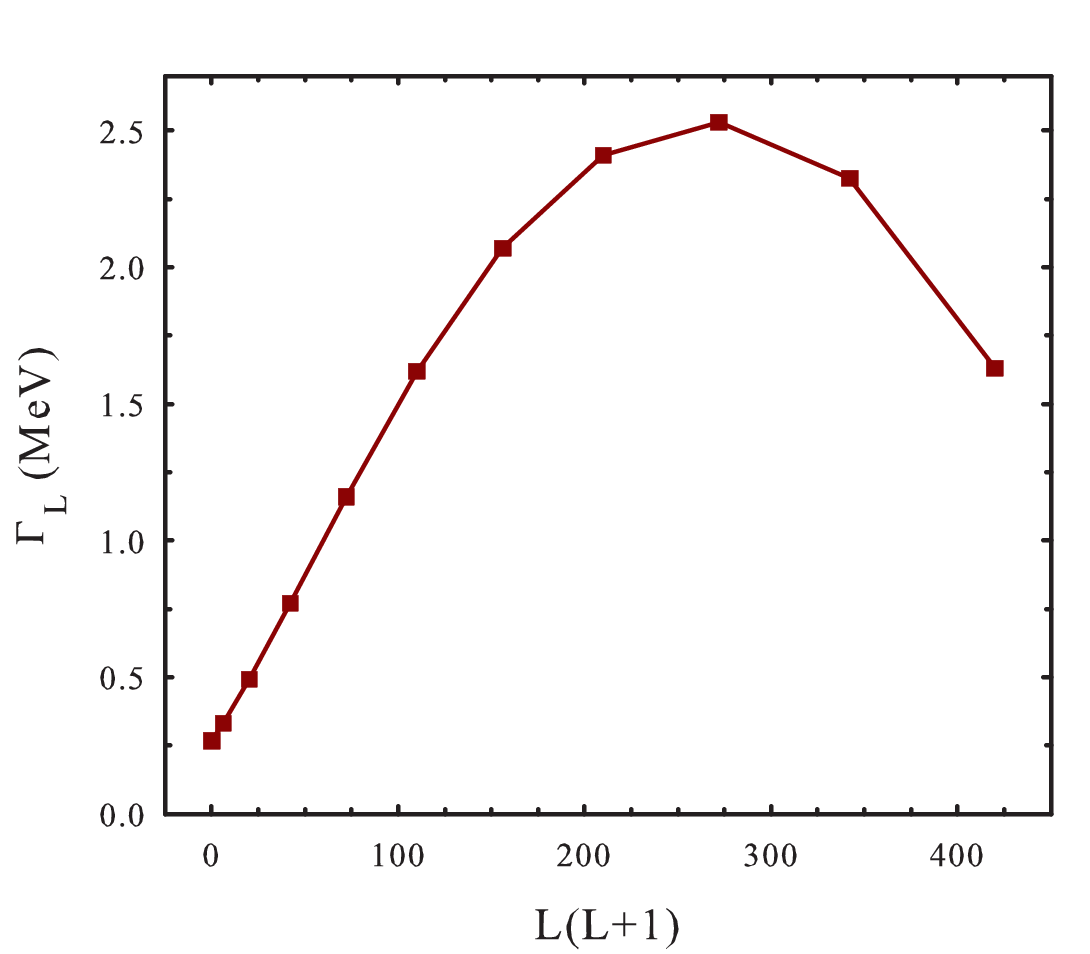}}
\caption{\small (Color online)
Total width of resonance states forming the first rotational band as a function of factor $L\left( L+1 \right)$.
\label{Fig:WidthResonsvsL}}
\end{figure}

In Fig. \ref{Fig:EnergyResonsvsL} we display energy of resonance states as a function of
$L\left(  L+1\right)  $. Indeed, the resonance states from 0$^{+}$ to 20$^{+}$
exhibit linear dependence on $L\left(  L+1\right)  $ and thus form two
rotational bands with a fixed moment of inertia. The first rotational band
involve all resonance states from 0$^{+},$ to 20$^{+}$, the second rotational
band is formed by very narrow resonance states with $L$=10, 12, 14, 16 \ and
18. Thus, two interacting nuclei \isotope[12]{C} behave as a rigid rotating body in
all detected resonance states. It is interesting that the total width of the
resonance states forming the first rotational band depend linearly on the
factor $L\left(  L+1\right)  $ for the orbital momentum $L\leq$10, as shown in
Fig.~\ref{Fig:WidthResonsvsL}.%

%


%

\begin{figure}[htbp]
\centerline{\includegraphics[width=88mm]{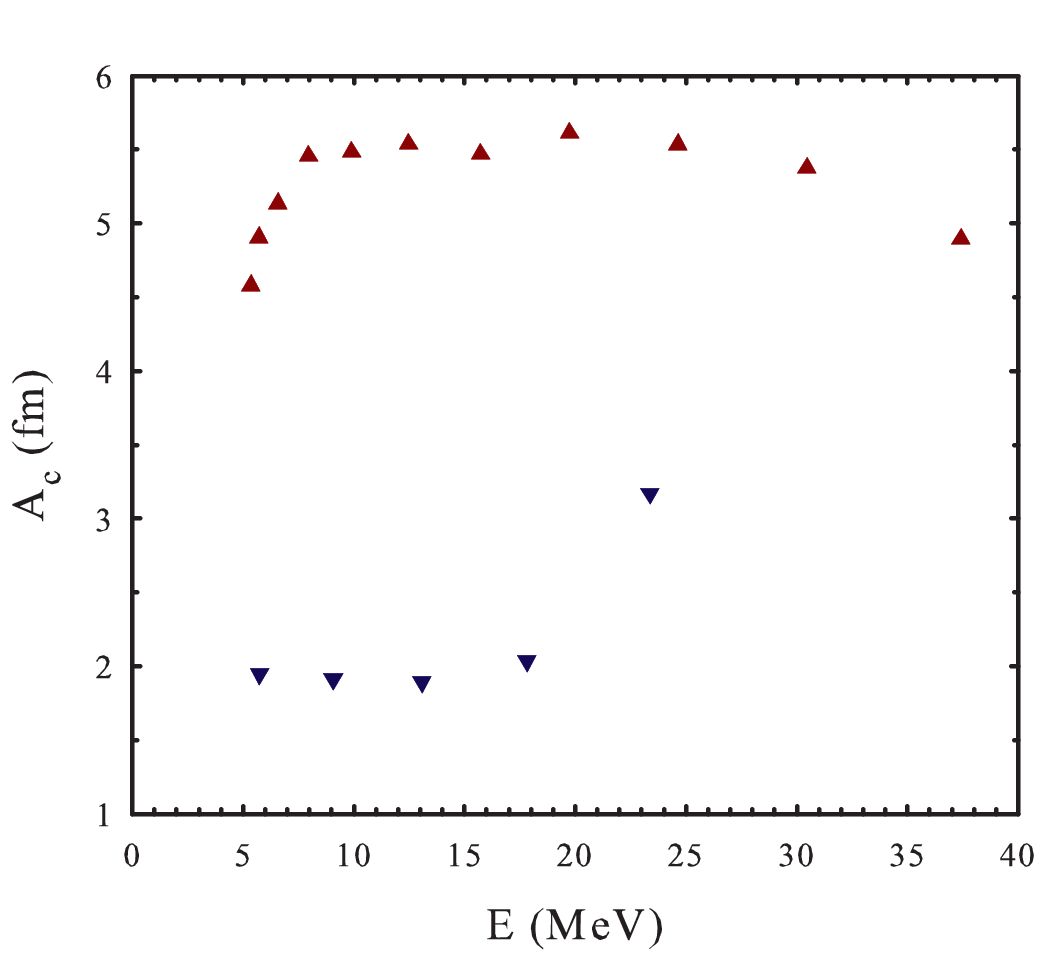}}
\caption{\small (Color online)
The average distance netween clusters as a function of the energy of resonance state.
\label{Fig:AverDistancvsE}}
\end{figure}

Let us consider correlation between the average distances between clusters and
energy of resonance state. This correlation is shown in Fig.~\ref{Fig:AverDistancvsE} where triangles up depicted average distances for
broad resonance states and triangles down represent set of very narrow
resonance states. Both set show quite regular correlations between average
distance and the energy of resonance states.%

%

\begin{figure}[htbp]
\centerline{\includegraphics[width=88mm]{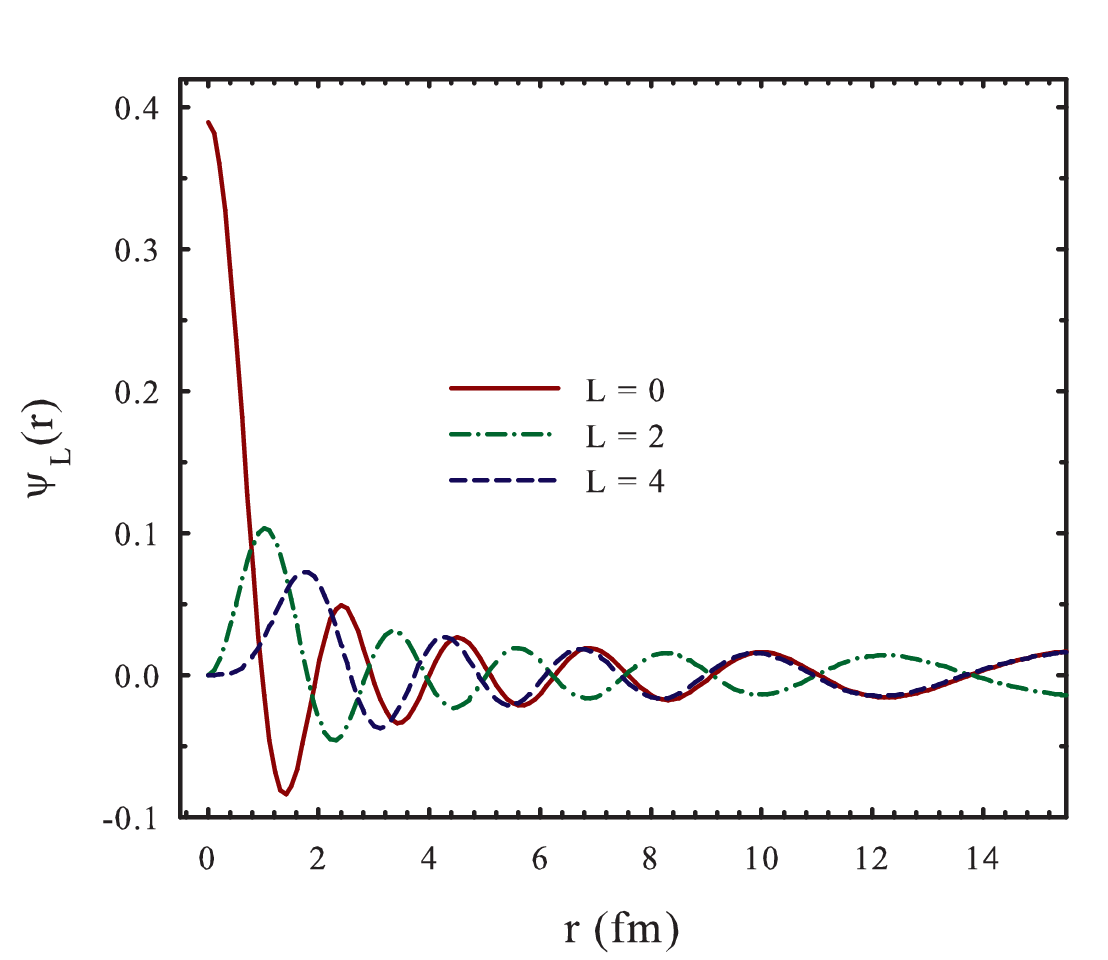}}
\caption{\small (Color online)
Wave functions of the $0^{+}$, $2^{+}$ and $4^{+}$ resonance states as a function of distance between center of mass of two \isotope[12]{C} nuclei\label{Fig:ResonWaveFunsFP}}
\end{figure}

Fig.~\ref{Fig:ResonWaveFunsFP} demonstrates behavior of wave functions for the
$0^{+}$, $2^{+}$ and $4^{+}$ resonance states determined with the folding
potential. One notices that the wave function of the $0^{+}$ resonance state
has very large value at $r$=0, wave functions of the $2^{+}$ and $4^{+}$ equal
zero at this point but they have relatively distinguished maxima at small
distance between two interaction \isotope[12]{C} nuclei.%
%

\begin{figure}[htbp]
\centerline{\includegraphics[width=88mm]{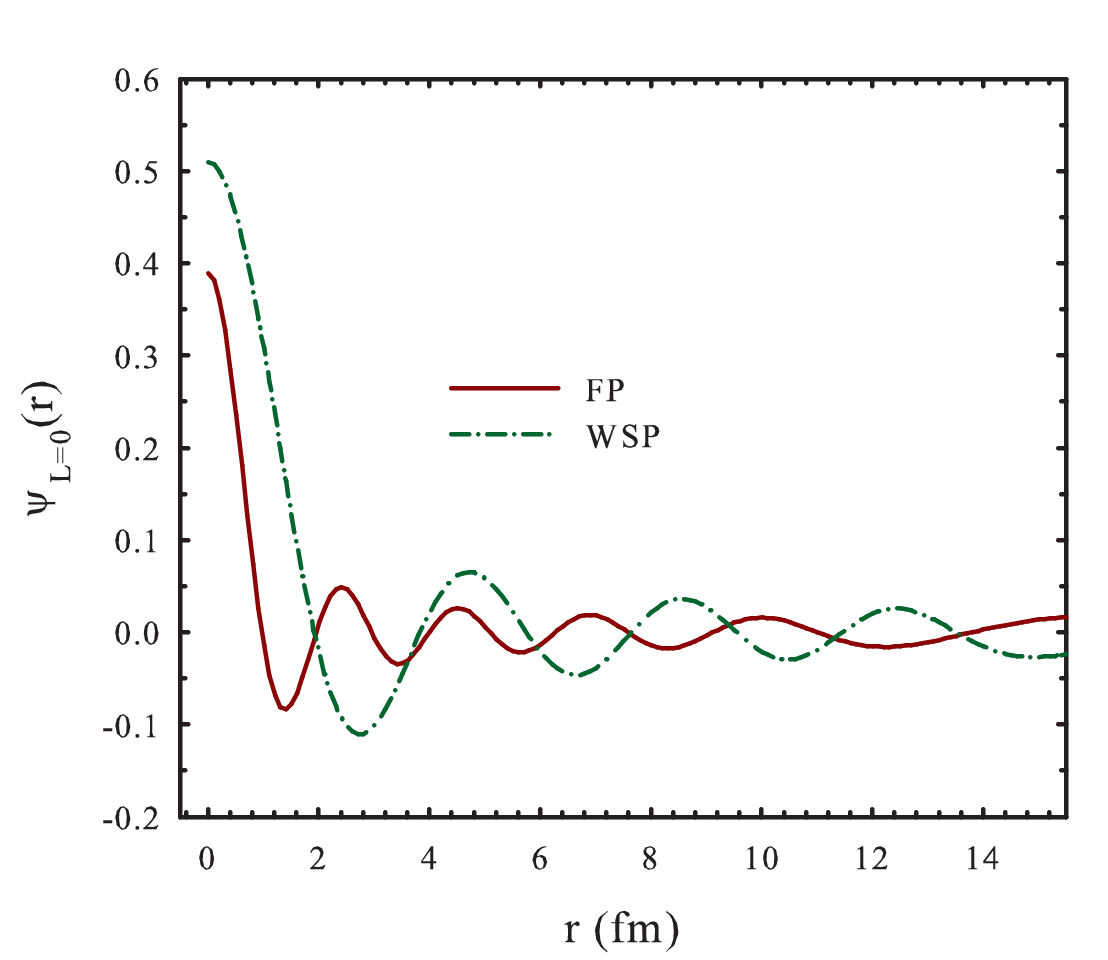}}
\caption{\small (Color online)
Wave functions of the $0^{+}$ resonance state obtained with the folding potential (FP) and the Woods-Saxon potential (WSP).
\label{Fig:Reson0PWaveFuns}}
\end{figure}

In Fig.~\ref{Fig:Reson0PWaveFuns} we compare the wave functions of the 0+ resonance states determined
with the folding potential and the Woods-Saxon potential. There
definite similarities of two wave function, both of them have large maximum at
$r = 0$, and oscillating behavior \ for $r > 0$.
The main difference between wave functions reflect that the folding
potential has a deeper potential well than the Woods-Saxon potential resulting
in larger frequency of oscillations.

%



Furthermore, we consider different ways to verify whether
parameters of resonance state are correctly determined. Within the present
model, we use few alternative methods. The first and the main method to
determine energy and width of a resonance state, is a standard way, it start
with calculations of phase shifts and the energy range where phase shifts are
relatively fast growing with increasing of the energy indicates possible
position of resonance state. By approximating the calculated phase shifts
$\delta$ with the Breit-Wigner formula
\begin{equation}
\delta=\delta_{p}-\arctan\left(  \frac{\Gamma}{E-E_{r}}\right)  ,
\label{eq:R08}%
\end{equation}
we obtain parameters of resonance state. Here, $\delta_{p}$ is the background
phase shift and the second term in Eq. (\ref{eq:R08}) is the resonance phase
shift. More specifically, we employ the following relations to determine
energy of resonance state $E_{R}$ and the total width $\Gamma$:
\begin{equation}
\left.  \frac{d^{2}\delta\left(  E\right)  }{dE^{2}}\right\vert _{E=E_{r}%
}=0,\quad\Gamma=2\left[  \frac{d\delta\left(  E\right)  }{dE}\right]
_{E=E_{r}}^{-1}. \label{eq:R010}%
\end{equation}
We assumed that the first derivative of the background phase shift with
respect to energy is much more smaller than the first derivative of the
resonance phase shift with respect to energy.

\begin{figure}[htbp]
\centerline{\includegraphics[width=88mm]{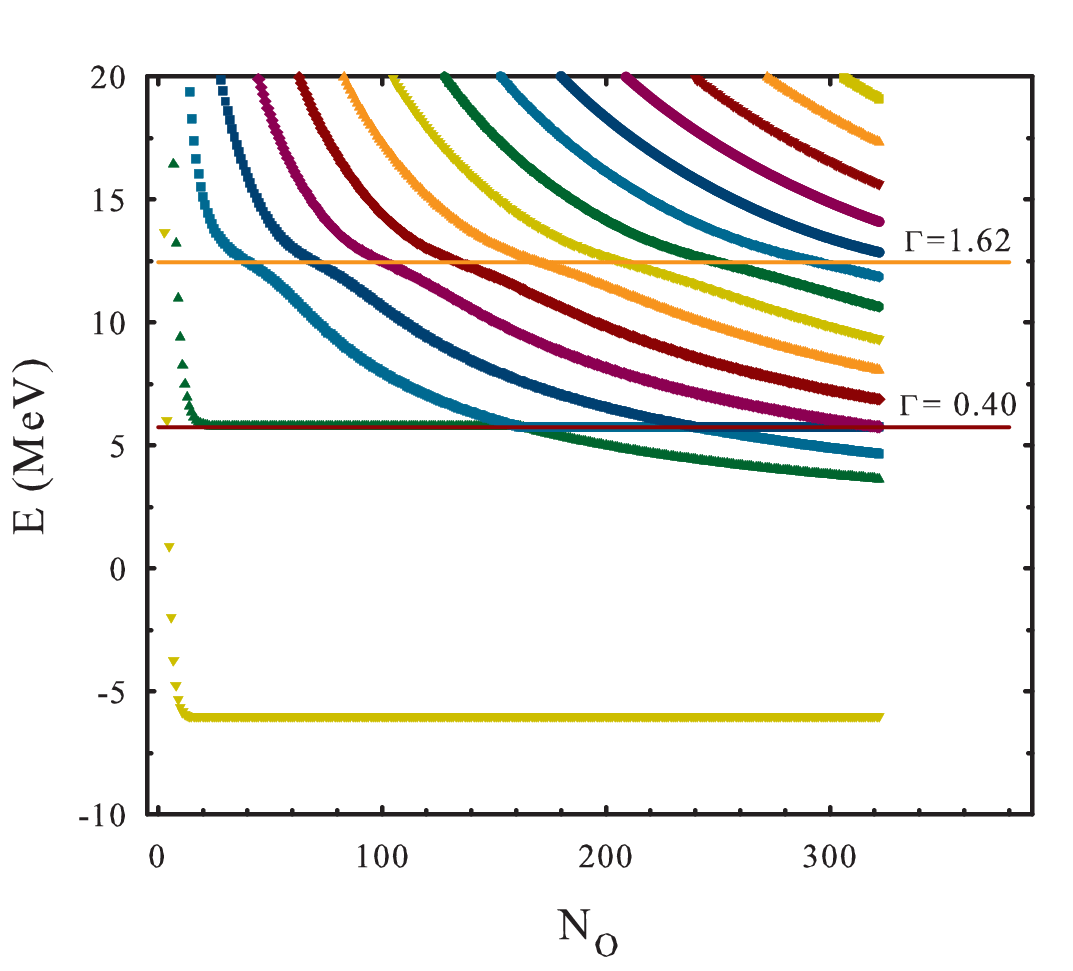}}
\caption{\small (Color online)
Spectrum of the eigenenergies of the 10$^{+}$ state in $^{24}$Mg as a function of the number oscillator functions $N_{O}$ involved in calculations
\label{Fig:Spectr24Mg10P}}
\end{figure}

There are another ways to confirm existence of resonance state and to find its
position. The first way is the stabilization method Ref.
\cite{1970PhRvA...1.1109H}. To realize it, one needs to calculate
eigenspectrum of the Hamiltonian as a function of the number of
square-integrable functions. In our method, we use the basis of oscillator
functions. Narrow resonance state manifest itself a plateau, as it
demonstrated in Fig. \ref{Fig:Spectr24Mg10P} where eigenspectrum for the
10$^{+}$ state are displayed. One can also see in Fig. \ref{Fig:Spectr24Mg10P} that to find a very narrow
resonance state we need at list 25 oscillator functions. Actually in this state we detected two resonance
states. It is worth while noticing that the almost same procedure is used to
locate resonance states in the Complex Scaling Method, details of formulation
and recent progress of the method can be found in Refs.
\cite{2020PTEP.2020lA101M, 2014PrPNP..79....1M, 2012PThPS.192....1H}.


%

\begin{figure}[htbp]
\centerline{\includegraphics[width=88mm]{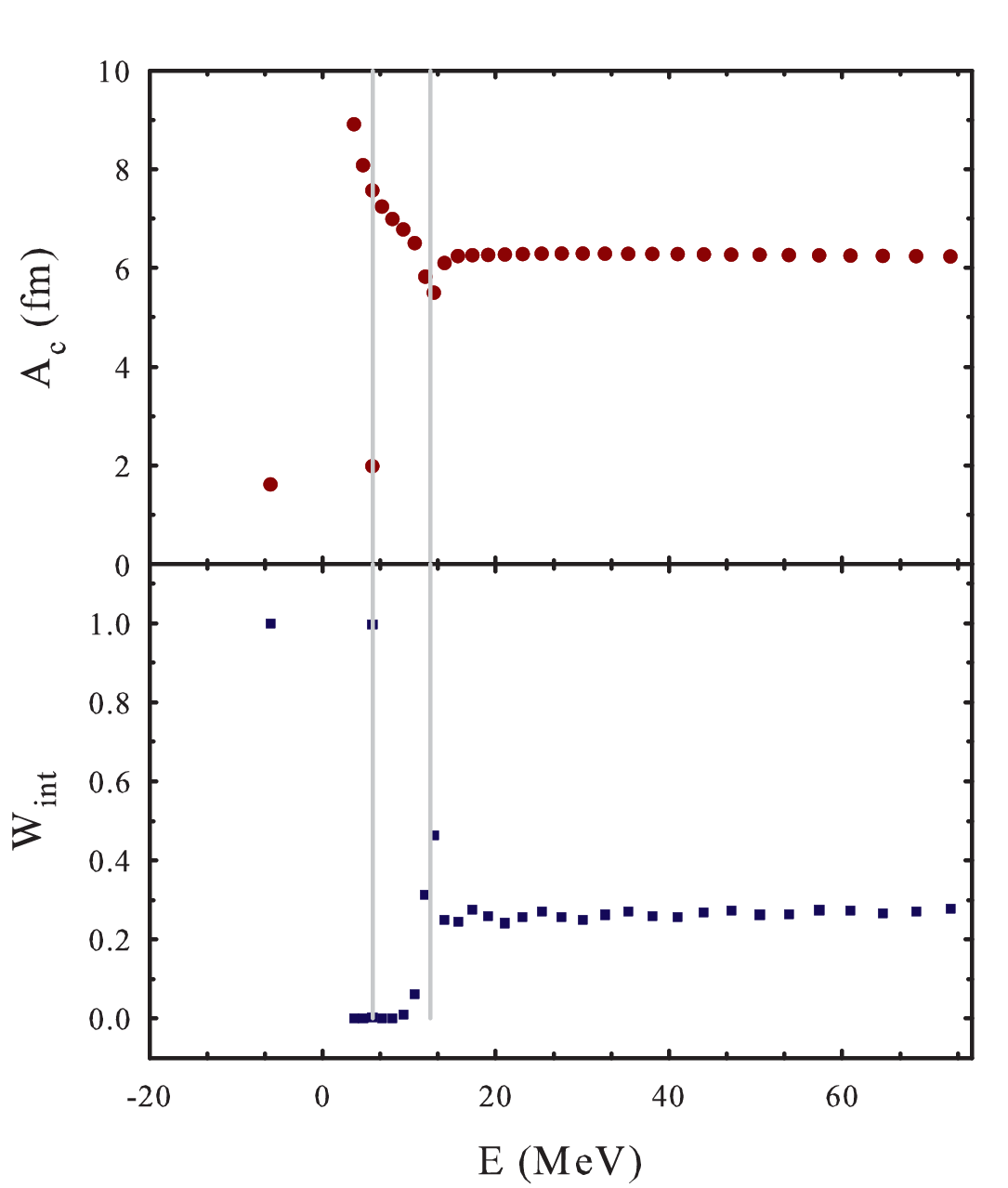}}
\caption{\small (Color online)
Average distance $A_{c}$ between two \isotope[12]{C} nuclei (the upper part) and weight \ $W_{int}$ of the internal part of wave function of the $10^+$ state as function of energy $E$.
\label{Fig:AverDistWint10P}}
\end{figure}

Another alternative method for detecting resonance state, is presented in Fig.
\ref{Fig:AverDistWint10P}\ where the average distance between clusters $A_{c}$
and the weight $W_{int}$ of the internal part of the wave function of
intercluster motion as a function of energy are depicted. The vertical lines
indicate the position of resonance states determined with the standard method.
We can see that the average distance between \isotope[12]{C} cluster in the narrow
resonance state is only 1.9 fm, as it was indicated in Table
\ref{Tab:ResonsVP}, while for wide resonance state it exceeds 5 fm. Both
values are smaller than average distances for other states displayed in Fig.
\ref{Fig:AverDistWint10P}. The internal weight for narrow resonance state
reach its maximum and is close to 100\%, the internal weight for wide
resonance state has a contribution of 46\% which is substantially larger than
for other states of continuous spectrum. Thus, such quantities can be used to
prove resonance states and locate their positions.

%






\section{Method of quantum mechanics for nucleus-nucleus scattering with fusion
at closed distances
\label{sec.3.2}}

In this Section we will study process of incident collision of two nuclei starting from very small relative distance 
(typical for distances between the closest nuclei located the lattice sites in neutron star) 
with possibility of their fusion.
This process can be studied on the basis of solution of Schr\"{o}dinger equation with radial potential
which has barrier approximated by large number $N$ of rectangular steps
%
%
\begin{equation}
  V(r) = \left\{
  \begin{array}{cll}
    V_{1}   & \mbox{at } r_{\rm min} < r \leq r_{1}      & \mbox{(region 1)}, \\
    \ldots   & \ldots & \ldots \\
    V_{N_{\rm fus}}   & \mbox{at } r_{N_{\rm fus} - 1} \leq r \leq r_{N_{\rm fus}}         & \mbox{(region $N_{\rm fus}$)}, \\
    \ldots   & \ldots & \ldots \\
    V_{N}   & \mbox{at } r_{N-1} \leq r \leq r_{\rm max} & \mbox{(region $N$)},
  \end{array} \right.
\label{eq.mir.1.1}
\end{equation}
where $V_{j}$ are constants ($j = 1 \ldots N$),
%
$r_1$ \ldots $r_N$ are parameters of discretization scheme with constant step used in computer calculations. One can calculate these parameters as follows:
\begin{equation}
\begin{array}{llllllll}
\vspace{0.5mm}
  \Delta r = \displaystyle\frac{r_{\rm max} - r_{\rm min}}{N}, \\

\vspace{0.5mm}
  r_{1} = \Delta r \cdot 1 + r_{\rm min}, &
  r_{N-1} = \Delta r \cdot (N-1) + r_{\rm min}, \\
\vspace{0.5mm}
  r_{2} = \Delta r \cdot 2 + r_{\rm min}, &

  r_{N} = \Delta r \cdot N + r_{\rm min} =
  r_{\rm max}. \\

\vspace{0.5mm}
  r_{i} = \Delta r \cdot i + r_{\rm min},
\end{array}
\label{eq.mir.1.2}
\end{equation}
We denote the first region with a left boundary at point $r_{\rm min}$ (we assume $r_{\rm min} \le 0$).
But, in addition to study in Ref.~\cite{Maydanyuk.2015.NPA},
now we assume that the fusion between nuclei takes place in region
$N_{\rm fus}$ after tunneling of particle with reduced mass through the barrier from the right part of potential, and
next propagation to the left part is possible and can be studied also.

A general solution of the radial wave function (up to its normalization) for the above barrier energies is
\begin{equation}
\chi(r) = \left\{
\begin{array}{lll}
   \alpha_{1}\, e^{ik_{1}r} + \beta_{1}\, e^{-ik_{1}r},
     &  r_{\rm min} < r \leq r_{1}, \\ 
   \alpha_{2}\, e^{ik_{2}r} + \beta_{2}\, e^{-ik_{2}r},
     &  r_{1} \leq r \leq r_{2}, \\ 
   \ldots & \ldots  \\
   \alpha_{N-1} e^{ik_{N-1}r} + \beta_{N-1} e^{-ik_{N-1}r}, &  r_{N-2} \leq r \leq r_{N-1}, \\ 
   e^{-ik_{N}r} + A_{R}\,e^{ik_{N}r}, & r_{N-1} \leq r \leq r_{\rm max}, 
\end{array} \right.
\label{eq.mir.1.3}
\end{equation}
where $\alpha_{j}$ and $\beta_{j}$ are unknown amplitudes, 
$A_{R}$ is unknown amplitude of the full reflection (from region of barrier and internal nuclear region),
and $k_{j} = \frac{1}{\hbar}\sqrt{2m(\tilde{E}-V_{j})}$ are complex wave numbers.
We fix the normalization so that the modulus of amplitude of the starting wave $e^{-ik_{N}r}$ equals to unity.

We will study this problem by the multiple internal reflections approach.
Note the important advance of this method as it can be applied for 
(1) study of scattering of nuclei at condition of experimental laboratories on Earth
(where we have nuclei in beam and in target), and
(2) nuclei incident on each other from very close distances, related with lattice sites in neutron stars 
(that is not studied by other known quantum methods).
This method was described in details in Ref.~\cite{Maydanyuk.2015.NPA} for capture of $\alpha$-particles by nuclei 
(see Refs.~\cite{Maydanyuk.2000.UPJ,Maydanyuk.2002.JPS,Maydanyuk.2011.JPS,Maydanyuk.2006.FPL,Maydanyuk_Zhang_Zou.2017.PRC} for other nuclear processes, time durations,
other aspects in scattering)
where we presented details of formalism, demonstrated its accuracy in comparison with other methods, we used tests to check calculations.
However, in Ref.~\cite{Maydanyuk.2015.NPA} it was not taken into account,
that after tunneling through the barrier
further propagation of waves inside the internal nuclear region of potential exists.
This logic follows from continuity of fluxes on the full region of definition of the full wave function of scattering, 
that is strict condition in quantum mechanics, it should be included to formalism.

At first, we calculate coefficients in each region of potential as
\begin{equation}
\begin{array}{ll}
\vspace{2mm}
   T_{j}^{+} = \displaystyle\frac{2k_{j}}{k_{j}+k_{j+1}} \,e^{i(k_{j}-k_{j+1}) r_{j}}, & 
   T_{j}^{-} = \displaystyle\frac{2k_{j+1}}{k_{j}+k_{j+1}} \,e^{i(k_{j}-k_{j+1}) r_{j}}, \\
   R_{j}^{+} = \displaystyle\frac{k_{j}-k_{j+1}}{k_{j}+k_{j+1}} \,e^{2ik_{j}r_{j}}, & 
   R_{j}^{-} = \displaystyle\frac{k_{j+1}-k_{j}}{k_{j}+k_{j+1}} \,e^{-2ik_{j+1}r_{j}}.
\end{array}
\label{eq.mir.1.4}
\end{equation}
Then, we find all summed amplitudes
$\tilde{R}_{N-2}^{+}$ \ldots $\tilde{R}_{N_{\rm fus}}^{+}$
and $\tilde{T}_{N-2}^{-}$ \ldots $\tilde{T}_{N_{\rm fus}}^{-}$
on the basis of recurrent relations
\begin{equation}
\begin{array}{llllllll}
  \tilde{T}_{j-1}^{-} =
    \displaystyle\frac{\tilde{T}_{j}^{-} T_{j-1}^{-}} {1 - R_{j-1}^{-} \tilde{R}_{j}^{+}}, \\

  \tilde{R}_{j-1}^{+} =
     R_{j-1}^{+} + \displaystyle\frac{T_{j-1}^{+} \tilde{R}_{j}^{+} T_{j-1}^{-}} {1 - \tilde{R}_{j}^{+} R_{j-1}^{-}}, \\

  \tilde{R}_{j+1}^{-} =
    R_{j+1}^{-} +
    \displaystyle\frac{T_{j+1}^{-} \tilde{R}_{j}^{-} T_{j+1}^{+}} {1 - R_{j+1}^{+} \tilde{R}_{j}^{-}},
\end{array}
\label{eq.mir.1.5}
\end{equation}
where for start we use
\begin{equation}
\begin{array}{cccc}
  \tilde{R}_{N-1}^{+} = R_{N-1}^{+}, & \quad
  \tilde{T}_{N-1}^{-} = T_{N-1}^{-}.
\end{array}
\label{eq.mir.1.6}
\end{equation}
On the basis of such amplitudes, we calculate summed amplitudes $\alpha_{j}$ and $\beta_{j}$ as
\begin{equation}
\begin{array}{lllllll}
  \beta_{j} \equiv
  \displaystyle\sum\limits_{i=1} \beta_{j}^{(i)} =
  \displaystyle\frac{\tilde{T}_{j}^{-}}{1 - \tilde{R}_{j-1}^{-} \tilde{R}_{j}^{+}}, \\

  \alpha_{j} \equiv
  \displaystyle\sum\limits_{i=1} \alpha_{j}^{(i)} =
  \displaystyle\frac{\tilde{R}_{j-1}^{-} \tilde{T}_{j}^{-}}{1 - \tilde{R}_{j-1}^{-} \tilde{R}_{1}^{+}}.
\end{array}
\label{eq.mir.1.7}
\end{equation}
%
%


\vspace{1.5mm}
\textbf{\emph{Summarised amplitudes of transition through the barrier and reflection from it.
\label{sec.mir.amplitudes.summarised}}}\;
\emph{Summarized amplitude $A_{T, {\rm bar}}$ of transition through the barrier} and
\emph{summarized amplitude $A_{R, {\rm bar}}$ of reflection from the barrier}
are determined as all waves transmitted through the potential region with the barrier from $r_{\rm fus}$ to $r_{N-1}$
or reflected from this potential region as
\begin{equation}
\begin{array}{llllll}
  A_{T, {\rm bar}} = \tilde{T}_{N_{\rm fus}}^{-}, &
  A_{R, {\rm bar}} = \tilde{R}_{N-1}^{-} &
  {\rm at}\;
  \tilde{R}_{N_{\rm fus}}^{-} = R_{N_{\rm fus}}^{-}.
\end{array}
\label{eq.mir.2.1}
\end{equation}
%

\vspace{1.5mm}
\textbf{\emph{Resonant and potential scatterings.
\label{sec.mir.scattering}}}\;
According to the method of multiple internal reflections,
\emph{potential scattering} is defined as the summarized amplitude $A_{R, {\rm ext}}$ of all waves reflected from region between 
the external barrier point $r_{\rm ext}$
(this can be external turning point $r_{\rm tp,ext}$ for under-barrier energies)
and $r_{N-1}$
(i.e., without inclusion of propagation through any part of the barrier region),
and propagated outside as
\begin{equation}
\begin{array}{llllll}
\vspace{1.5mm}
  S_{\rm pot} \equiv
  A_{R, {\rm ext}} = \tilde{R}_{N-1}^{-},
  {\rm at}\;
  \tilde{R}_{N_{\rm tp,ext}}^{-} = R_{N_{\rm tp,ext}}^{-}.
\end{array}
\label{eq.mir.2.2}
\end{equation}
\emph{Resonant scattering} is defined as 
the summarized amplitude $A_{R, {\rm tun}}$ of all waves which are reflected from the potential region between 
point $r_{\rm fus}$ and 
the external barrier point $r_{\rm ext}$
(this can be external turning point $r_{\rm tp,ext}$ for under-barrier energies)
as
\begin{equation}
\begin{array}{lll}
  S_{\rm res} \equiv
  A_{R, {\rm tun}} = A_{R, {\rm bar}} - A_{R, {\rm ext}}, &
  S = S_{\rm res} + S_{\rm pot}.
\end{array}
\label{eq.mir.2.3}
\end{equation}
Here, $S$ is the full amplitude of $S$-matrix of scattering.
%

\vspace{1.5mm}
\textbf{\emph{Coefficients of penetrability, reflections, oscillations.
\label{sec.mir.penetrability}}}\;
Coefficients of penetrability $T_{\rm bar}$ and 
reflection $R_{\rm bar}$ concerning potential region,
coefficient $R_{\rm ext}$ of reflection from the external part of the barrier
coefficient $R_{\rm tun}$ of reflection from the barrier region
are 
\begin{equation}
\begin{array}{lllll}
\vspace{1.5mm}
   T_{\rm bar} = \displaystyle\frac{k_{\rm fus}}{k_{N}}\; \bigl\|A_{T, {\rm bar}} \bigr\|^{2}, &
   R_{\rm bar} = \bigl\|A_{R, {\rm bar}}\bigr\|^{2}, \\
   R_{\rm ext} = \bigl\|A_{R, {\rm ext}}\bigr\|^{2}, &
   R_{\rm tun} = \bigl\| A_{R, {\rm tun}}\bigr\|^{2}.
\end{array}
\label{eq.mir.3.1}
\end{equation}
\emph{Amplitude of oscillations} and \emph{coefficient of oscillations}
concerning to point of fusion with number $N_{\rm fus}$ are
\begin{equation}
\begin{array}{lllllll}
\vspace{1.5mm}
  A_{\rm osc} = 
  \displaystyle\frac{1}{1 - \tilde{R}_{N_{\rm fus}-1}^{-} \tilde{R}_{N_{\rm fus}}^{+}}, & 
  K_{\rm osc} = \bigl\| A_{\rm osc} 
  \bigr\|^{2}.
\end{array}
\label{eq.mir.3.2}
\end{equation}
Standard test of quantum mechanics, $T_{\rm bar} + R_{\rm bar} = 1$, is naturally used in this formalism,
in order to check calculations.
Important to note that this test has not been used in other approaches to check calculations of fusion and scattering processes in nuclear physics,
so this is advance of our formalism.


\vspace{1.5mm}
\textbf{\emph{Probability of existence of the compound nucleus.
\label{sec.mir.compoundnucleus}}}\;
During collision of the incident nucleus \isotope[12]{C} with another nucleus \isotope[12]{C} located in the lattice site 
a compound nucleus can be formed from these two nuclei.
Probability of existence of this compound nucleus is defined as integral over region 
between two internal turning points 
(these points indicate internal space region before barrier on the radial semi-axis for under-barrier energies,
see Refs.~\cite{Maydanyuk_Zhang_Zou.2017.PRC,Maydanyuk_Shaulskyi.2022.EPJA}).

In case, if there is no fusion processes in the compound nucleus, 
the probability of its existence is written down as
[see Eq.~(17) in Ref.~\cite{Maydanyuk_Zhang_Zou.2017.PRC}]
%
\begin{equation}
\begin{array}{llllll}
  P_{\rm cn}^{\rm (without\, fusion)}  \equiv 
  \displaystyle\int\limits_{r_{\rm int,1}}^{r_{\rm int,2}} \|\chi(r)\|^{2}\; dr =

  \displaystyle\sum\limits_{j=1}^{n_{\rm int}}
  \Bigl\{
    \bigl( \|\alpha_{j}\|^{2}  \\

  + \;\;
    \|\beta_{j}\|^{2} \bigr)\, \Delta r +
    \displaystyle\frac{\alpha_{j}\beta_{j}^{*}} {2ik_{j}}\,  e^{2ik_{j}r}
      \Bigr\|_{r_{j-1}}^{r_{j}} -
    \displaystyle\frac{\alpha_{j}^{*}\beta_{j}} {2ik_{j}}\,  e^{-2ik_{j}r}
      \Bigr\|_{r_{j-1}}^{r_{j}}
  \Bigr\}.
\end{array}
\label{eq.mir.4.1}
\end{equation}
%
Solution above is essentially simplified for the simplest barrier 
which can be presented only by two potential regions in Eqs.~(\ref{eq.mir.1.1}) 
(see Eqs.~(1), (6), (7) in Ref.~\cite{Maydanyuk_Zhang_Zou.2017.PRC}):
\begin{equation}
\begin{array}{llllll}
  P_{\rm cn}^{\rm (without\, fusion)}  =
    P_{\rm osc}\, T_{\rm bar}\, P_{\rm loc}, \\

  P_{\rm osc} = \|A_{\rm osc}\|^{2} \\

  \hspace{6.0mm} = 
  \displaystyle\frac{(k + k_{1})^{2}}
    {2k^{2} (1 -\cos (2k_{1}r_{1})) + 2k_{1}^{2}\,(1 + \cos (2k_{1}r_{1})) }, & \\

  T_{\rm bar} \equiv \displaystyle\frac{k_{1}}{k_{2}}\; \bigl\| T_{1}^{-} \bigr\|^{2}, 
  \\

  P_{\rm loc} = 2\, \displaystyle\frac{k_{2}}{k_{1}}\; \Bigl( r_{1} - \displaystyle\frac{\sin(2k_{1}r_{1})}{2k_{1}} \Bigr).
\end{array}
\label{eq.mir.4.2}
\end{equation}
%
%
%
%
%
For fast fusion with simple barrier we obtain ($R_{0} \equiv 0$)
\begin{equation}
\begin{array}{lllll}
  P_{\rm cn}^{\rm (fast\, fusion)} =
  \Bigl\| \displaystyle\sum\limits_{i=1} \beta_{1}^{(i)} \Bigr\|^{2}
  \displaystyle\int\limits_{0}^{r_{1}}
    \Bigl\| R_{0} e^{ik_{1}r} + e^{-ik_{1}r} \Bigr\|^{2}\, dr \\
%

  \;\;\; = \;
  \bigl\| T_{1}^{-} \bigr\|^{2}\, r_{1} =
  \displaystyle\frac{k_{2}\, r_{1}}{k_{1}} T_{\rm bar}, \\

   T_{1}^{-} = \displaystyle\sum\limits_{i=1} \beta_{1}^{(i)}, 
   \hspace{10.0mm}
   T_{1}^{-} = \displaystyle\frac{2k}{k+k_{1}}\,e^{-i(k-k_{1})r_{1}}.
\end{array}
\label{eq.mir.4.3}
\end{equation}
%
Note that condition $R_{0} \equiv 0$ used in this formula
corresponds to \emph{the sharp angular momentum cutoff} 
(see Eq.~(3), Ref.~\cite{Maydanyuk.2015.NPA})
introduced by Glas and Mosel in Refs.~\cite{Glas.1975.NPA,Glas.1974.PRC}.
Many researchers used this condition in calculations of cross sections of fusion (also captures by nuclei),
while our formula (\ref{eq.mir.4.3}) estimates probability of the compound nucleus formation 
with complete fast fusion and further formation of new nucleus \isotope[24]{Mg}.


\vspace{1.5mm}
\textbf{\emph{Cross section of fusion.
\label{sec.mir.crosssection.fusion}}}\;
%
%
The fusion cross section is defined as
(see Ref.~\cite{Maydanyuk.2015.NPA}, for details):
%
%
\begin{equation}
\begin{array}{lll}
  \sigma_{\rm fus}^{\rm (not\, fast)} (E) = \displaystyle\sum\limits_{l=0}^{+\infty} \sigma_{l}(E), \\
  \sigma_{l} = \displaystyle\frac{\pi\hbar^{2}}{2mE}\, (2l+1)\, f_{l}(E)\, P_{\rm cn} (E).
\end{array}
\label{eq.mir.5.1}
\end{equation}
%
%
Here,
$E$ is the energy of the relative motion between two nuclei,
$\sigma_{l}$ is the partial cross-section at $l$,
$P_{\rm cn}$ is probability of formation of compound nuclear system defined in Eqs.~(\ref{eq.mir.4.1})--(\ref{eq.mir.4.3}).
%
%
Connecting the old factor of fusion $P_{l}$ with new probability $P_{\rm cn} (E)$ and penetrability of barrier region $T_{{\rm bar,} l}(E)$
for the fast fusion,
coefficient $f_{l}(E)$ is calculated as
%
\begin{equation}
  f_{l}^{\rm (fast)} (E) = \displaystyle\frac{k_{\rm fus}}{k_{N}\, \|r_{\rm fus} - r_{\rm tp,in, 1}\|}.
\label{eq.mir.5.2}
\end{equation}
To describe formation of the compound nucleus with slow fusion (i.e., not fast fusion),
we vary fusion coefficients in the region between points $r_{\rm fus}$ and $r_{\rm int, 2}$.

We define also \emph{cross section of fast fusion} as
\begin{equation}
\begin{array}{lll}
  \sigma_{\rm fus}^{\rm (fast)} (E) = \displaystyle\sum\limits_{l=0}^{+\infty} \sigma_{l}(E), \\
  \sigma_{l}(E) = \displaystyle\frac{\pi\hbar^{2}}{2mE}\, (2l+1)\, T_{{\rm bar,} l}\, P_{l}^{\rm (fast)},
\end{array}
\label{eq.mir.5.3}
\end{equation}
where fusion probability is 
\begin{equation}
  P_{l}^{\rm (fast)} =
  \left\{
  \begin{array}{lll}
    1 & \mbox{\rm at } & l = 0, \\
    0 & \mbox{\rm at } & l > 1.
  \end{array}
\right.
\label{eq.mir.5.4}
\end{equation}
%





\vspace{-7mm}

\section{Resonances in formation of compound nuclei in reaction: 
folding approach with S-form
\label{sec.analysis.9}}

\vspace{-2.0mm}
We study the $\isotope[12]{C} + \isotope[12]{C}$ collision
with folding potential of $S$-form 
starting from relative distance between nuclei located in lattice sited of neutron star.
Numerical results show that penetrability of the barrier of folding potential increases
and reflection decreases monotonously with
increasing of energy of the incident nucleus
(
see Fig.~\ref{fig.fold.shellP.1},
this property of penetrability and reflection is observed up to 150 MeV,
this is in agreement with results in Ref.~\cite{Maydanyuk_Shaulskyi.2022.EPJA}).
Thus, penetrability and reflection do not form resonant states of compound nucleus during collision
(this was also stated for capture of $\alpha$ particles by nuclei~\cite{Maydanyuk.2015.NPA,Maydanyuk_Zhang_Zou.2017.PRC}).

The probability of formation of compound nucleus in $\isotope[12]{C} + \isotope[12]{C}$
calculated by the method of Multiple Internal Reflections 
is shown in Fig.~\ref{fig.fold.shellP.2}~(a).
The maxima in the probability function are clearly observed, indicating on new states in which a compound nucleus is formed in pycnonuclear reactions with the highest probability.
These maxima are explained by strict requirement of quantum mechanics of conservation of the full flux inside the full region of definition of the wave function. 
This condition requires to take into account the further propagation of quantum fluxes in the nuclear region,
in contrast to the modern description of pychonuclear reactions, 
where these fluxes are ignored in the nuclear region. 
%
%
Thus, synthesis of more heavy nucleus \isotope[24]{Mg} after fusion at energies of such states is much more probable 
than at energies of zero mode vibrations
in neutron stars~\cite{ShapiroTeukolsky.2004.book,Zeldovich.1965.AstrJ}

Coefficient of resonant scattering has clear maxima 
(see Fig.~\ref{fig.fold.shellP.3})
while penetrability has monotonous behavior in dependence on energy 
(so, it does not give maxima in calculation of coefficient of resonant scattering). 
Presence of such maxima and sharp minima in coefficient of resonant scattering is explained by influence from oscillations of fluxes
in the internal nuclear region.
Now, if to look at the coefficient of oscillations, one can see maxima and minima in this coefficient
[see Fig.~\ref{fig.fold.shellP.2}~(b)].


Energies of quasibound states for $\isotope[12]{C} + \isotope[12]{C}$ up to 150~MeV
are given in Tabl.~\ref{table.fold.1}.
where we also add results for 
the Woods-Saxon potential~\cite{Maydanyuk.2023.Universe}.
It truns out that the energies of the quasibound states are different significantly depending on the potential type.
That indicates 
importance of accurate determining the potential of interaction between nuclei
in study of synthesis in stars.

\begin{table}
\hspace{-20mm}
\begin{tabular}{|c|c|c|c|c|c|c|c|c|c|} \hline
  No. & WS-form & 
  $S$-form &
  $F$-form
  \\ \hline

  1   &  4.881 &  2.492 &  3.487 \\
  2   & 11.450 & 10.452 & 8.462 \\
  3   & 20.408 & 22.641 &  16.422 \\
  4   & 31.456 & 38.312 &  26.621 \\
  5   & 43.699 & 56.968 &  38.063 \\
  6   &  57.136 & 78.111 &  50.998 \\
  7   &  71.767 & 101.245 &  65.176 \\
  8   &  87.294 & 126.866 &  80.350 \\
  9   & 104.016 &  & 96.519 \\
  10  & 121.931 &  & 113.931 \\



\hline
\end{tabular}
\caption{Energies (in MeV) of the quasibound states of the compound nucleus in fusion
$\isotope[12]{C} + \isotope[12]{C}$
calculated by the method of multiple internal reflections with the different potentials 
up to 150~MeV.
Here, Coulomb 2 (WS-form) is calculation for potential of Woods-Saxon type 
in Eqs.~(\ref{eq.potentialWS.2a})--(\ref{eq.potentialWS.2}),
Coulomb 3 ($S$-form) is a new calculation for the folding approach
(nuclear part of potential is determined in for $S$-form in 
Eq.~(\ref{eq.fold.nuclei-s.potential.C12})
with parameter $z_{\nu}$ given in Eq.~(\ref{eq:R022}),
for simplicity Coulomb part of potential is defined in Eqs.~(\ref{eq.potentialWS.2}),
Coulomb 4 ($F$-form) is a new calculation for the folding approach with F-form
[such a potential is defined in Eqs.~(\ref{eq:R025})--(\ref{eq:R025})]. 
Accuracy about $10^{-14}$ in checking test $|T_{\rm bar} + R_{\rm bar}| = 1$ is obtained in all calculations.
Only first quasibound energies for $\isotope[12]{C} + \isotope[12]{C}$ are smaller than barrier maximums for these nuclear systems.
At such energies the compound nuclear systems have barriers which prevent decays going through tunneling.
}
\label{table.fold.1}
\end{table}

In Tabl.~\ref{table.fold.2} one can see that
energies of maxima of the coefficients of the probability of formation of the compound nucleus do not coincide with energies of the maxima of the oscillation coefficient. 
Thus, knowledge of the resonance scattering is not enough to find maxima of the probability of formation of the compound nucleus. 
%
\begin{table}
\begin{center}
\begin{tabular}{|c|c|c|c|c|c|c|c|c|} \hline


 No. & Comp. nucleus 
     & Resonant scattering,  
     & Oscillations, \\

     & formation, $P_{cn}$
     & $S_{\rm res.\, scat.}$ 
     & $T_{\rm osc.}$ \\ \hline


 1 & 2.4924 &
 1.9949 & 
 2.4924 \\


 2 & 10.4524 & 
 3.7362 & 
 12.1936 \\


 3 & 22.6410 & 
 29.6060 & 
 26.6210 \\


 4 & 38.3121 & 
 43.2871 & 
 37.5659 \\


 5 & 56.9682 & 
 62.6894 & 
 59.2070 \\


 6 & 78.1118 & 
 85.0767 & 
 84.3305 \\


 7 & 101.2454 & 
 109.9515 & 
 112.6878 \\


 8 & 126.8664 & 
 136.3188 & 
 143.2838 \\

\hline
\end{tabular}
\end{center}
\vspace{-1.0mm}
\caption{Energies (in MeV)
of quasibound states of the compound nucleus, 
maxima of the resonant scattering coefficient, 
maxima of the oscillation coefficient 
for $\isotope[12]{C} + \isotope[12]{C}$. 
Energies of the quasibound states of compound nucleus and the maxima of the coefficients do not coincide.
}
\label{table.fold.2}
\end{table}

Note that the accuracy of 
the method of Multiple Internal Reflections in calculation of these coefficients is $10^{-14}$ or higher~\cite{Maydanyuk.2015.NPA,Maydanyuk.2023.Universe}. The sub-barrier and above-barrier energies are also included to this analysis.



\vspace{-1.5mm}
\section{Resonances in formation of compound nuclei in reactions: folding approach with F-form
\label{sec.analysis.shells_p}}

\begin{figure}[htbp]
\centerline{\includegraphics[width=88mm]{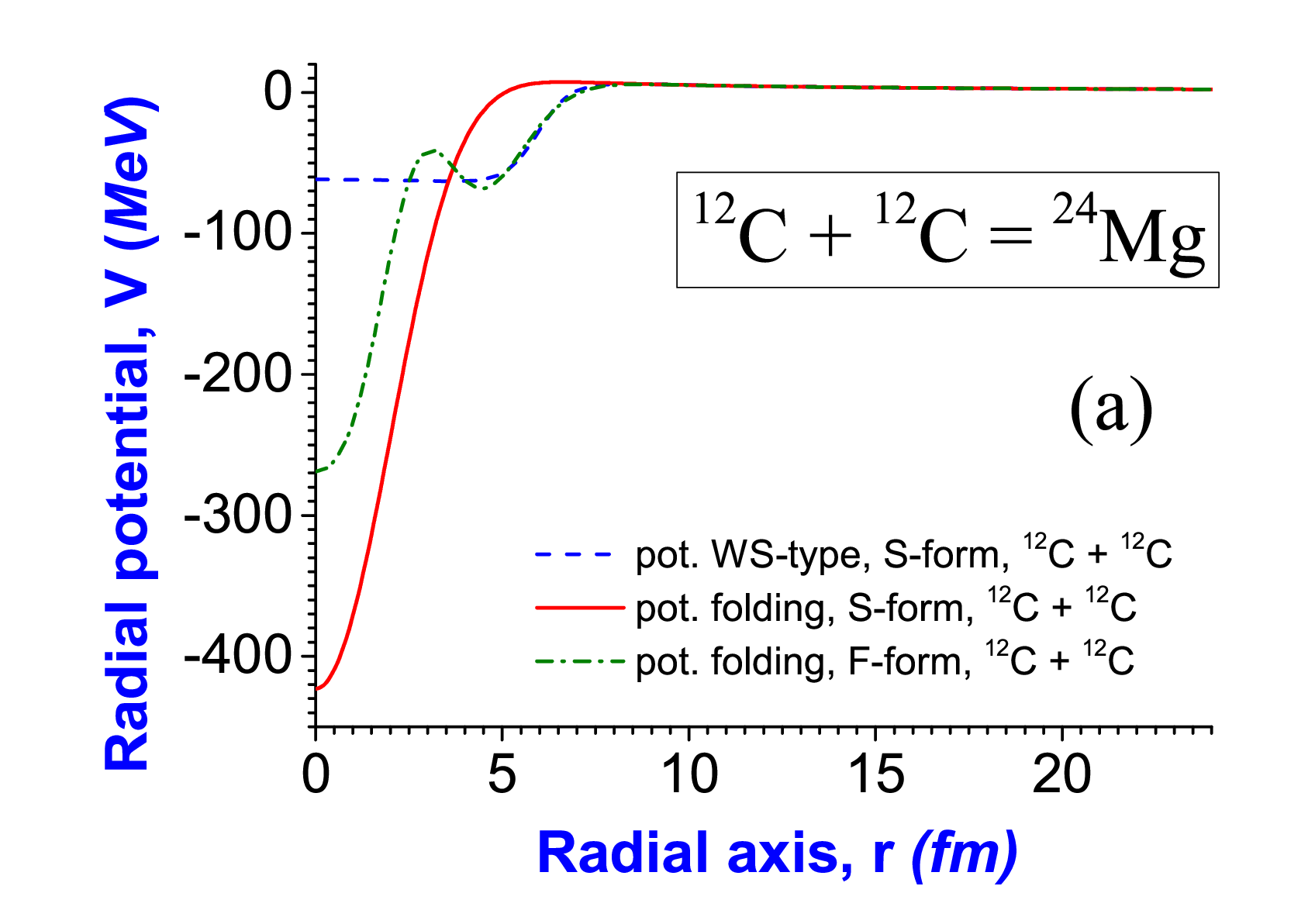}}
\centerline{\includegraphics[width=88mm]{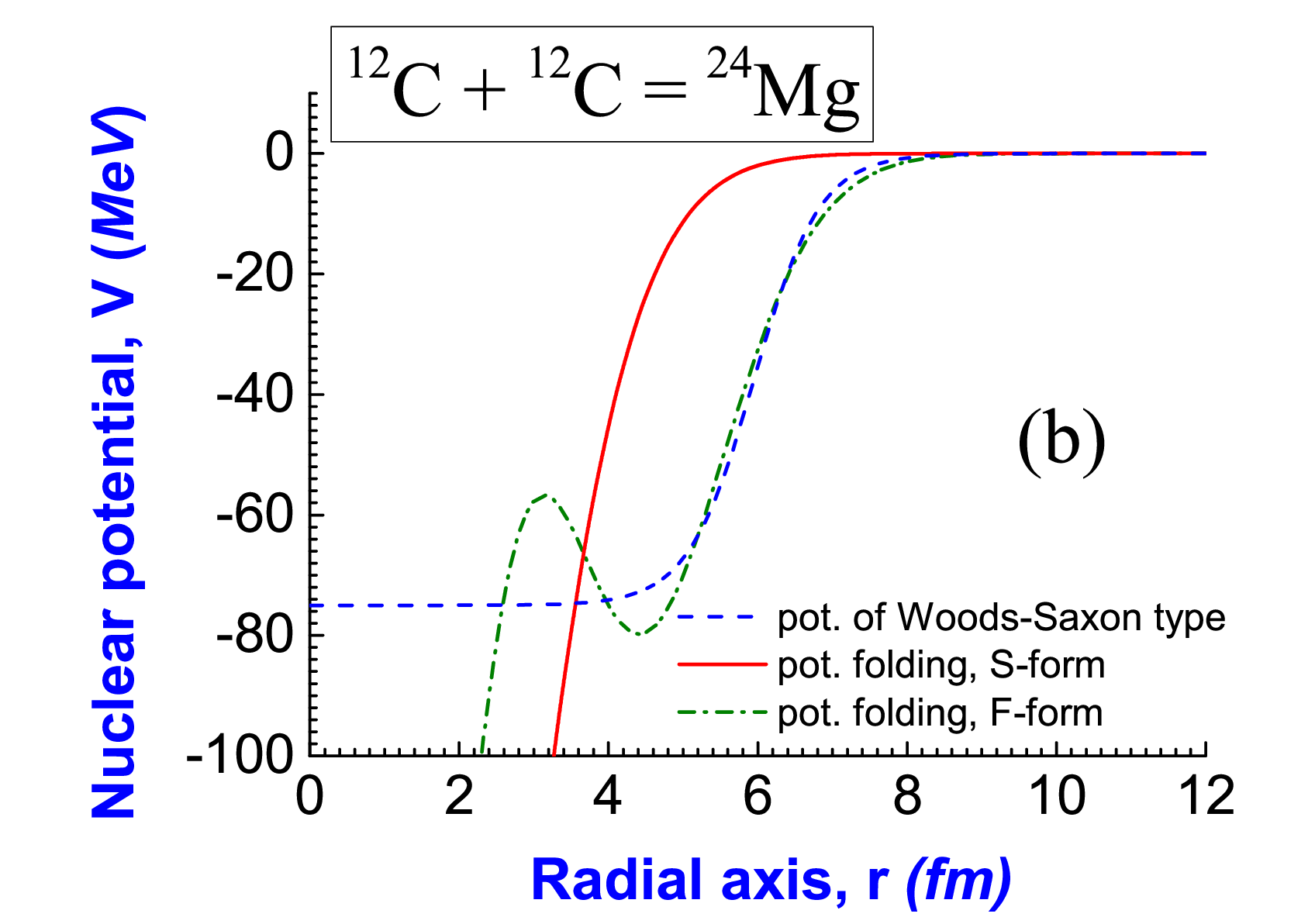}}
\vspace{-3mm}
\caption{\small (Color online)
The folding potential with F-form in comparison with
the potential of Woods-Saxon type and 
the folding potential with S-form.
Note that new folding potential with F-form has two barriers
(that phenomenon is very unusual in physics of nuclear reactions).
The full potentials (a) and nuclear parts of these potentials (b) are shown.
\label{fig.fold.potential_P.1}}
\end{figure}

\vspace{-1.5mm}
Now we will study process of formation of compound nucleus during collision $\isotope[12]{C} + \isotope[12]{C}$ by the folding approach with F-form.
%
%
The potential for $\isotope[12]{C} + \isotope[12]{C}$ 
calculated in the folding approach with F-form is shown in Fig.~\ref{fig.fold.potential_P.1} 
in comparison with 
the potential of the Woods-Saxon type and
the potentials calculated on the basis of the folding approach with S-form.
Note that such a new potential has the second smaller barrier
($V_{\rm bar,\, 1} = -56.6$~MeV at $r = 3.12$~fm).
This has not been observed before in nuclear physics for tasks with potentials of scattering.
It is explained by taking into account $F$-form in calculation of the nuclear part of potential 
[see Eqs.~(\ref{eq:R025})].

\begin{figure}[htbp]
\centerline{\includegraphics[width=88mm]{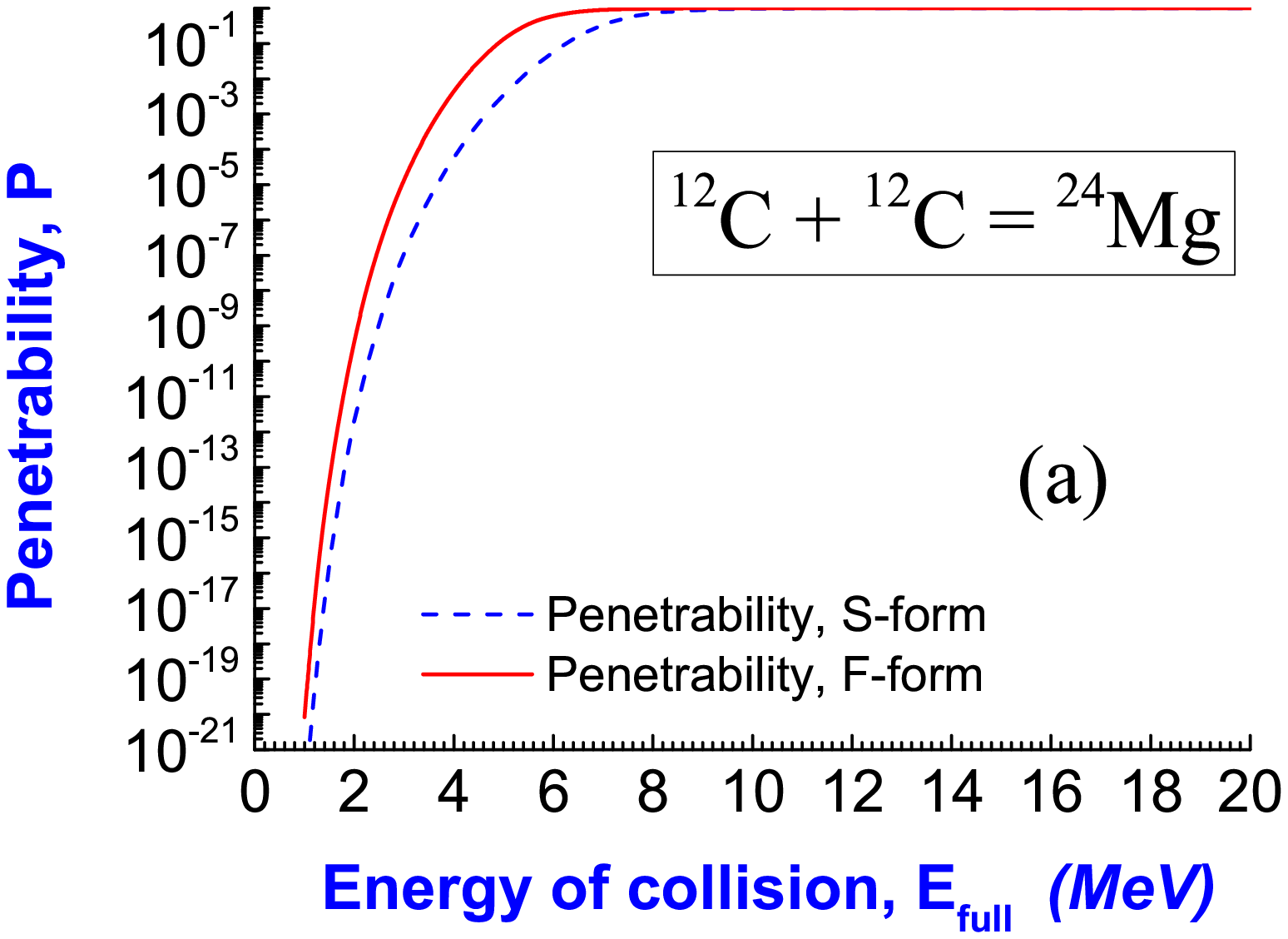}}
\centerline{\includegraphics[width=88mm]{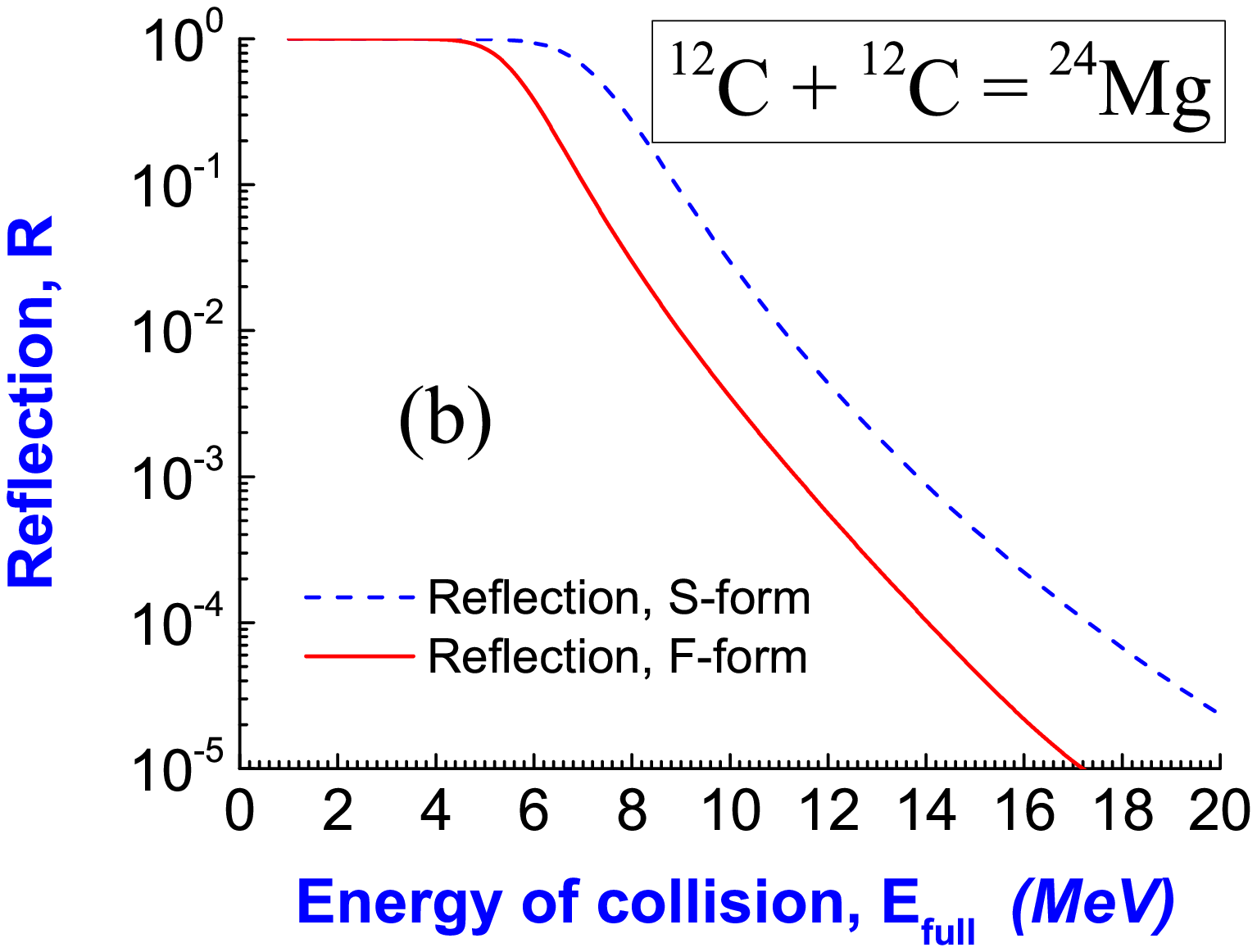}}
\vspace{-3mm}
\caption{\small (Color online)
Coefficients of penetrability $T_{\rm bar}$ of barrier region (a) and reflection $R_{\rm bar}$ from it (b) in dependence on 
energy of collision 
(energy of relative motion between the incident nucleus and nucleus in lattice cite) 
for $\isotope[12]{C} + \isotope[12]{C}$ 
with the nucleus-nucleus potential derived by the folding approach.
%
%
%
%
\label{fig.fold.shellP.1}}
\end{figure}

Analysis has shown that penetrability of barrier region for the nucleus-nucleus potential 
increases
and reflection decreases monotonously with
increasing of energy of collision (see Fig.~\ref{fig.fold.shellP.1}).
%
%
%
One can see that there is no any maxima in such curves,
so, the penetrability and reflection do not form any resonant states in collision.
In case of F-form, penetrability for the barrier region is larger than for case of S-form.
Therefore, after taking into account case of $F$-form (that is more accurate calculation),
nucleus becomes more transparent during collision 
in comparison with case of S-form.

\begin{figure}[htbp]
\centerline{\includegraphics[width=88mm]{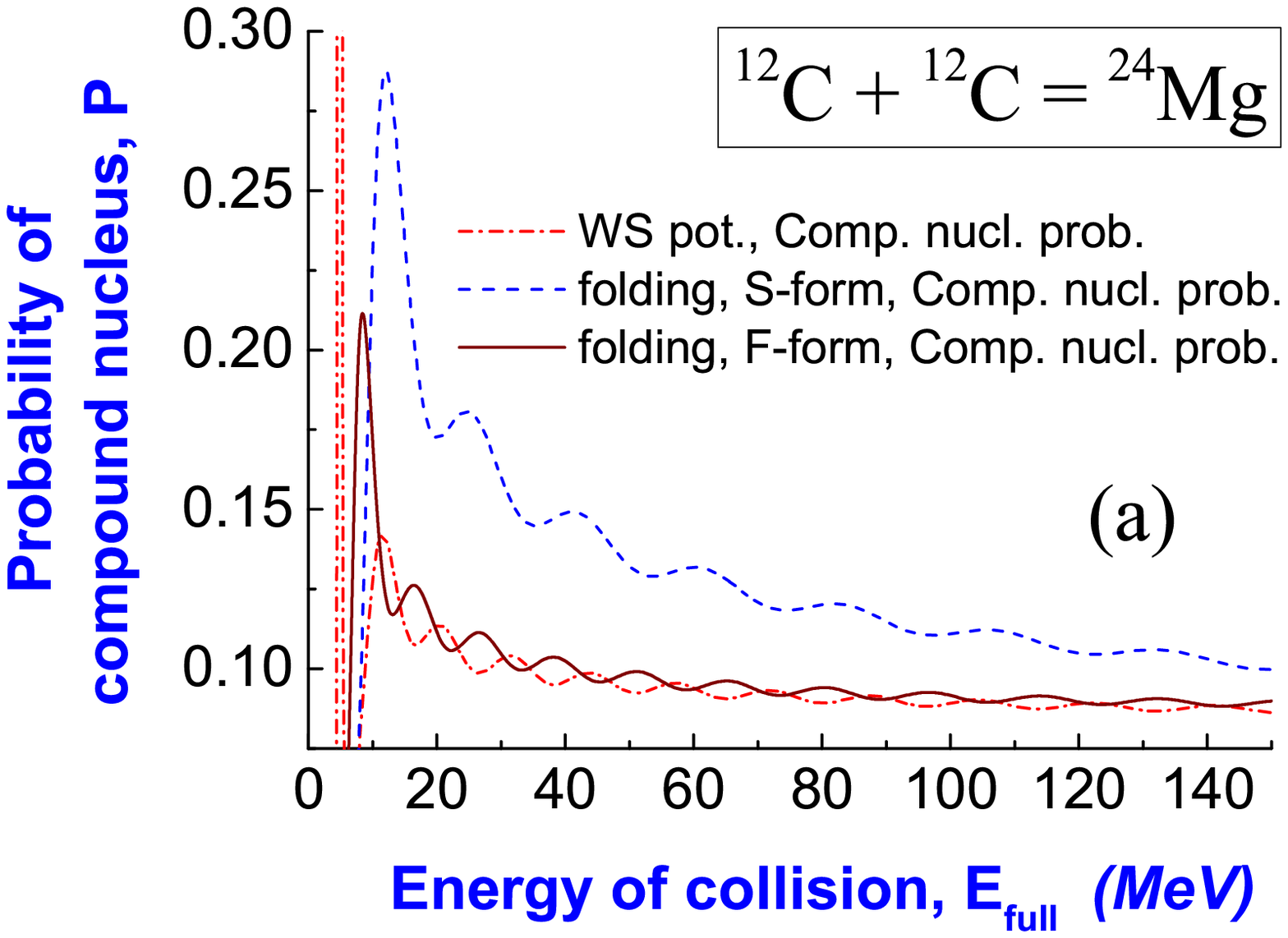}}
\centerline{\includegraphics[width=88mm]{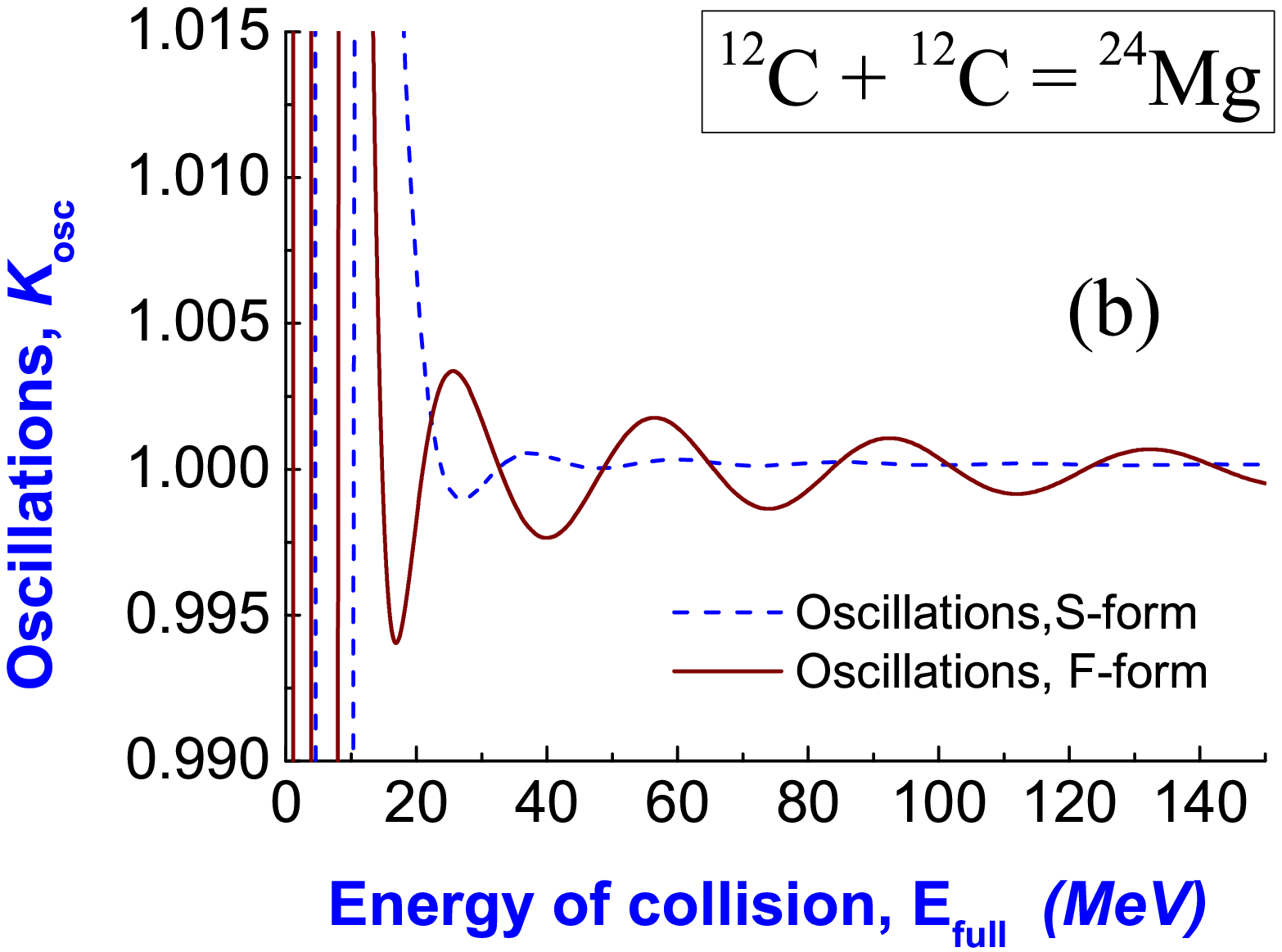}}
\vspace{-3mm}
\caption{\small (Color online)
Probabilities of formation of compound nucleus $P_{\rm cn}$ (a) and coefficient of oscillations $K_{\rm osc}$ (b)
in dependence on energy of collision 
in reaction $\isotope[12]{C} + \isotope[12]{C}$
calculated concerning to the nucleus-nucleus potentials in the folding approach
with case of F-form (see brown solid line).
%
\label{fig.fold.shellP.2}}
\end{figure}

\begin{figure}[htbp]
\centerline{\includegraphics[width=88mm]{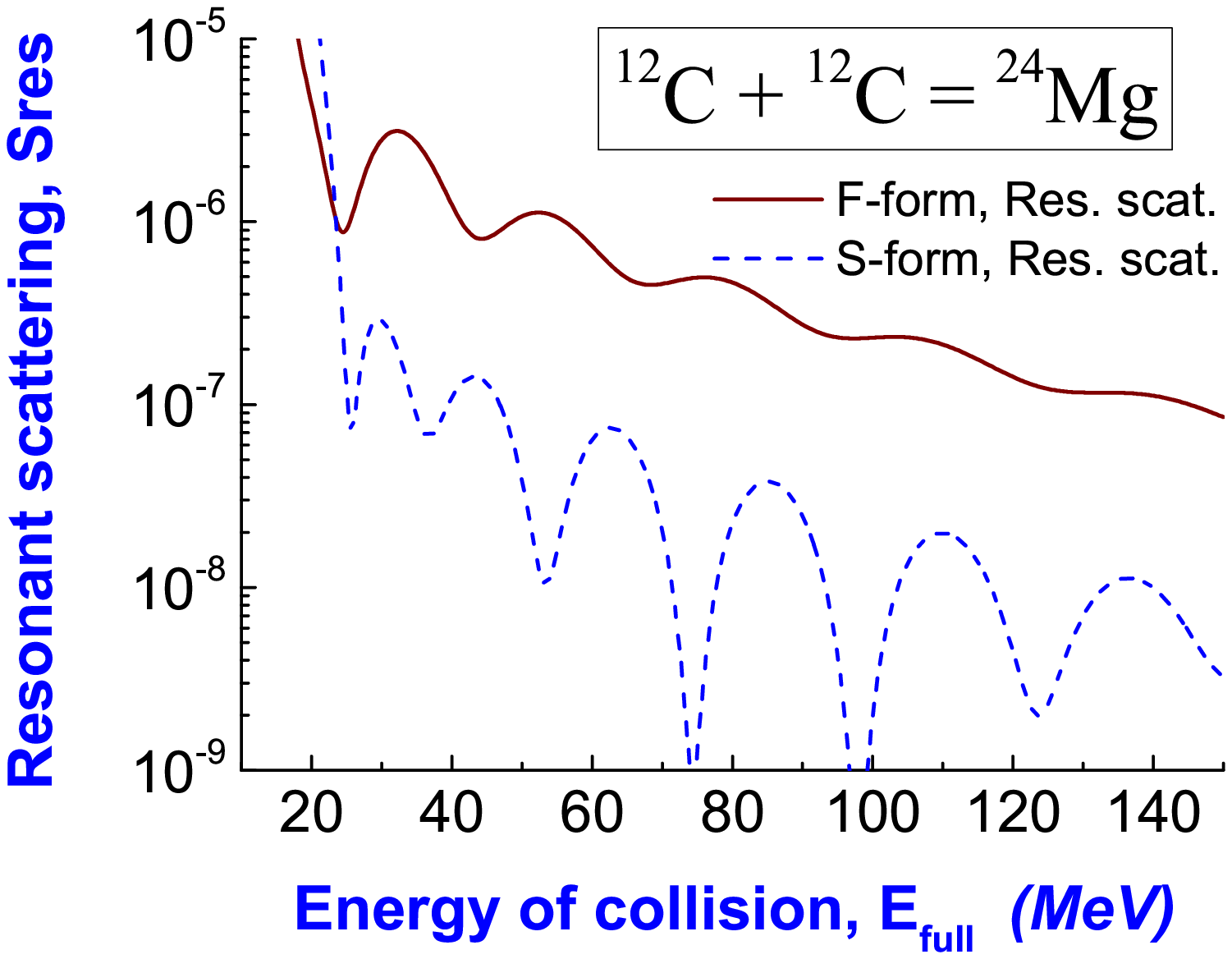}}
\vspace{-4mm}
\caption{\small (Color online)
Resonant scattering $S_{\rm res}$ 
in dependence on energy of collision 
in reaction $\isotope[12]{C} + \isotope[12]{C}$
calculated concerning to the nucleus-nucleus potentials in the folding approach
with cases of $S$-form (see blues dashed line) and $F$-form (see brown solid line).
%
%
\label{fig.fold.shellP.3}}
\end{figure}

The probability of compound nucleus formed during reaction $\isotope[12]{C} + \isotope[12]{C}$ and 
calculated concerning to the folding potential with F-form is presented in Fig.~\ref{fig.fold.shellP.2}~(a).
%
One can see presence of resonant maxima in the probability of formation of compound nucleus at certain energies for the folding potential with F-form.
However, these maxima in case of F-form are shifted to smaller energies (to the left part in figure) in comparison with the maxima in case of S-form.
In general, we find that modification of picture of formation of compound nucleus with these resonant energies 
calculated on the basis of the folding approach with case of F-form is essential
in comparison with old result obtained concerning to the potential of Woods-Saxon type in Ref.~\cite{Maydanyuk.2023.Universe}.
Note that energies of these maxima indicate on states where a compound nucleus is formed with the highest probability and 
synthesis of heavier nucleus \isotope[24]{Mg} after fusion is the most probable. 
This is a new result in understanding of pycnonuclear reactions in compact stars.
This result has confirmed importance of developments of microscopic approaches in study of physics of neutron stars.
%
In Tabl.~\ref{table.fold.1} we included energies of quasibound states for $\isotope[12]{C} + \isotope[12]{C}$ up to 150~MeV
concerning to the folding potential with F-form.
The energies of the quasibound states are changed significantly depending on the choice of the potential.
In Fig.~\ref{fig.fold.shellP.3} we add coefficient of resonant scattering 
calculated for the folding potential with $F$-form.
We obtained clear maxima in this coefficient.
Energies of such maxima are shifted to the larger values (to the right part in figure) in comparison with
results for the folding potential with case of S-form.
As for S-form, presence of such maxima 
in the resonant scattering is explained by influence from oscillations of fluxes
inside the internal nuclear region
[see coefficient of oscillations in Fig.~\ref{fig.fold.shellP.2}~(b)].


\vspace{-2.0mm}
\section{Conclusions and perspectives
\label{sec.conclusion}}

\vspace{-2.0mm}

We investigated possibility of synthesis of nuclei in pycnonuclear reactions in dense medium of neutron stars.
It is supposed that the compound nucleus is formed in collision of nuclei 
needed for synthesis of more heavy nucleus in dense stellar medium. Formation of the compound nucleus is studied with high precision on the basis of the method of Multiple Internal Reflections with tests and high accuracy of calculations.
%

Conclusions from our study are the following:

\vspace{-1.5mm}
\begin{itemize}
\item
New results of determination of potential of interactions between nuclei at close distances are obtained in the fully microscopic approach in condition of dense medium of neutron stars.
These are the folding approach with approximations of $S$-form, and new formalism of $F$-form. The folding approach with $S$-form was introduced in Refs.~\cite{Maydanyuk_Vasilevsky.2023.PRC,Shaulskyi_Maydanyuk_Vasilevsky.2024.PRC} 
for systematic study bremsstrahlung and reactions with light nuclei. The folding approach with $F$-form has never been presented before, and it is the most accurate approach in description of interactions between isotopes of Carbon today. Both approaches with $S$-form and $F$-form are applied to explore nuclear reactions in dense stellar medium of neutron stars.  


\vspace{-1.5mm}
\item
The inter-cluster folding potential formed by a nucleon-nucleon potential with coordinate part consisted of several Gaussians has been deduced. It is established that semi-realistic Volkov, Hasegawa-Nagata and Minnesota
potentials widely used in cluster models, 
create deep
attractive$\ ^{12}$C+$^{12}$C folding potentials of the different depth but
approximately of the same range. 
A set of resonance states were found for the orbital momentum $L$ from 0 to 20. 
It was established that the composition of the attractive nuclear potential and
centrifugal barrier combined with the Coulomb repulsive interaction creates
favorable conditions for residing of two resonance states, one is
narrow of order of several keV and another is wider than 1 MeV.  
Two resonance states were found for the orbital momenta from 10  to 18. Narrow and
wide resonance states split on two rotational bands.
Interacting nuclei $^{12}$C behave as a rigid
rotating body in all detected resonance states.
New algorithms were used to detected resonance states.
Mass root-mean-square radii of $^{24}$Mg and the average distances between two
$^{12}$C clusters in bound and resonance states of $^{24}$Mg were evaluated. 
Distance between clusters is less than 2 fm for narrow resonance states, for wide resonance states it varies from 4.6 to
5.6 fm.

\item
The folding potential with F-form has two barriers (see Fig.~\ref{fig.fold.potential_P.1}).
The second barrier ($V_{\rm bar,\, 1} = -56.6$~MeV at $r = 3.12$~fm)
has not been observed before in nuclear physics.
It is caused by inclusion of $p$-shell in calculation of potential 
[see Eqs.~(\ref{eq:R025})].

\item
For the process of collision of two nuclei from relative distance typical for close nuclei located in the lattice sites in neutron star,
we establish clear maxima of probability of formation of compound nucleus at some energies.
These energies correspond to states which we call as 
\emph{quasi-bound states in pycnonuclear reactions}~\cite{Maydanyuk.2023.Universe,Maydanyuk_Shaulskyi.2022.EPJA,Maydanyuk_Zhang_Zou.2017.PRC}.
At such energies the compound nucleus is formed with the highest probability and it is the most long lived.
Note that these states have not been predicted by other methods for study of synthesis of nuclei in stars.
%
We calculated these states for 
(1) potential of Woods-Saxon type [see Eqs.~(\ref{eq.potentialWS.2a})--(\ref{eq.potentialWS.2})],
(2) folding potential with 
S-form [see Eqs.~(\ref{eq.fold.nuclei-s.potential.C12})
%
],
(3) folding potential with 
F-form 
[see Eqs.~(\ref{eq:R025})--(\ref{eq:R025})].
Estimations show essential difference between quasibound energies for different potentials 
[see 
Fig.~\ref{fig.fold.shellP.2}, Tabl.~\ref{table.fold.1}].
So, cluster approach with the folding potential 
(see Fig.~\ref{fig.fold.shellP.2})
provides significant modification of the picture of formation of compound nucleus,
than
potential of Woods-Saxon type~\cite{Maydanyuk.2023.Universe}.

%

\vspace{-1.5mm}
\item
Formation of compound nucleus in the quasibound states is much more probable than in states of zero-point vibrations.

\vspace{-1.5mm}
\item
The first quasibound energies for
$\isotope[12]{C} + \isotope[12]{C}$ are smaller than the barrier maximums for all types of potentials.
We obtain
the first and second quasibound energies $E_{1} = 4.881$ and $E_{2} = 11.45 090$ for the potential of Woods-Saxon type,
the first and second quasibound energies $E_{1} = 2.492$ and $E_{2} = 10.452$ for the folding potential with S-form,
the first and second quasibound energies $E_{1} = 3.487$ and $E_{2} = 8.462$ for the folding potential with F-form
(see Tabl.~\ref{table.fold.1}).
The barrier maxima are
$V_{max} = 5.972$~MeV for the potential of Woods-Saxon type (at $r = 8.33$~fm),
$V_{max} = 7.366$~MeV for the folding potential with S-form (at $r = 6.65$~fm),
$V_{max} = 5.744$~MeV for the folding potential with F-form (at $r = 8.65$~fm)
(see Fig.~\ref{fig.fold.potential.1} and Fig.~\ref{fig.fold.potential_P.1}).
%
So, at the first energy 
the compound nuclear system has barrier which prevents decay 
going through tunneling.
Such a nuclear system represents bound system, this is the new 
nucleus \isotope[24]{Mg}
synthesized in dense medium of neutron star.

\vspace{-1.5mm}
\item
We study the colliding nuclei from close distances (about 50--200~fm) 
with the formation of compound nucleus 
and the subsequent synthesis of heavier nucleus in dense medium, which is assumed in a neutron star.
%
%
Other approaches describe collisions of nuclei starting from far asymptotic distances
that is related to experimental facilities with beams. 
In these approaches, pictures of formation of the compound nucleus are much different. 
So, 
development of methods of quantum mechanics with high precision 
for reactions at close distances
has good perspective. 
%
%
We determine the potential of interaction between nuclei in the fully microscopic approach based on two-nucleon interactions well tested on the available experimental data.
This approach allows to determine the structure of light nuclei, which is confirmed by experiments with high accuracy
\cite{Lashko_Vasilevsky_Zhaba.2024.PRC}.
It also provides safe basis to describe interactions between nuclei at close distances and in condition of dense medium of neutron star.
\end{itemize}

\vspace{-6.5mm}
\section*{Acknowledgements
\label{sec.acknowledgements}}

\vspace{-2.5mm}
Authors are highly appreciated to
Prof.~A.~G.~Magner for fruitful discussions concerning to nuclear matter in neutron stars and Earth.
%
%
%
This work is partly supported by the National Key R\&D Program of China under Grant No. 2023YFA1606703, and by the National Natural Science Foundation of China under Grant Nos. 12435007 and 12361141819. It is received partial support from the Program of Fundamental Research of the Physics and Astronomy Department of the National Academy of Sciences of Ukraine (Project No. 0122U000889). 
V.V.S. extends his gratitude to the Simons Foundation for financial support (Award ID: SFI-PD-Ukraine-00014580).

\end{document}